\documentclass[aps,prd,twocolumn,preprintnumbers,nofootinbib,amsmath,amssymb]{revtex4-2}
\usepackage{graphicx}
\usepackage{dcolumn}
\usepackage[pdftex, pdftitle={Continuum limit of the mobility edge and
  taste-degeneracy effects in high-temperature lattice QCD with
  staggered quarks},hidelinks]{hyperref}
\usepackage{epstopdf}
\usepackage{placeins}
\usepackage{multirow}
\usepackage{makecell}
\usepackage{cleveref}
\usepackage{xcolor}

\newcommand{\beq}{\begin{eqnarray}}
\newcommand{\eeq}{\end{eqnarray}}
\newcommand{\beqnn}{\begin{eqnarray*}}
\newcommand{\eeqnn}{\end{eqnarray*}}

\newcommand{\SU}{\mathrm{SU}}
\newcommand{\IPR}{\mathrm{IPR}}
\newcommand{\PR}{\mathrm{PR}}
\newcommand{\Var}{\mathrm{Var}}

\newcommand{\f}[2]{\frac{#1}{#2}}

\newcommand{\la}{\langle}

\newcommand{\ra}{\rangle}

\renewcommand{\Re}{{\rm Re}\,}
\renewcommand{\Im}{{\rm Im}\,}

\newcommand{\Oc}{{\cal O}}

\usepackage[font=small, labelfont=bf, textfont={small,it}]{caption}
\begin{document}

\title{Continuum limit of the mobility edge and taste-degeneracy
  effects in high-temperature lattice QCD with staggered quarks}

\author{Claudio Bonanno}
\email{claudio.bonanno@csic.es}
\affiliation{Instituto de F\'isica Te\'orica UAM-CSIC, c/ Nicol\'as
  Cabrera 13-15, Universidad Aut\'onoma de Madrid, Cantoblanco,
  E-28049 Madrid, Spain}

\author{Matteo Giordano}
\email{giordano@bodri.elte.hu}
\affiliation{ELTE E\"otv\"os Lor\'and University, Institute for
  Theoretical Physics, P\'azm\'any P\'eter s\'et\'any 1/A, H-1117,
  Budapest, Hungary}

\date{\today}

\begin{abstract}
  We study the effects of taste degeneracy on the continuum scaling of
  the localization properties of the staggered Dirac operator in
  high-temperature QCD using numerical simulations on the lattice,
  focusing in particular on the position of the mobility edge
  separating localized and delocalized modes at the low end of the
  spectrum.  We find that, if the continuum limit is approached at
  fixed spatial volume, the restoration of taste symmetry leads to
  sizeable systematic effects on estimates for the mobility edge
  obtained from spectral statistics, which become larger and larger as
  the lattice spacing is decreased.  Such systematics, however, are
  found to decrease if the volume is increased at fixed lattice
  spacing.  
  We argue that spectral statistics estimate correctly the position of
  the mobility edge in the thermodynamic limit at fixed spacing, and
  support this with an independent numerical analysis based directly
  on the properties of the Dirac eigenvectors, that are unaffected by
  taste degeneracy.
  We then provide a theoretical argument justifying the exchange of
  the thermodynamic and continuum limits when studying
  localization. 
  This allows us to use spectral statistics to determine the position
  of the mobility edge, and to obtain a controlled continuum
  extrapolation of the mobility edge for the first time. 
\end{abstract}

\maketitle

\section{Introduction}
\label{sec:intro}

The microscopic mechanism behind the finite-temperature transition in
QCD is still the subject of intense research activity. One of the open
questions is how the approximate restoration of chiral symmetry and
the deconfinement of quarks and gluons into the quark-gluon plasma,
both taking place as a rapid crossover in the same range of
temperatures~\cite{Borsanyi:2010bp,Bazavov:2016uvm}, are related to
each other.

An intriguing aspect of the transition, revealed by first-principles
nonperturbative numerical studies on the lattice, is that it is
accompanied by a radical change in the localization properties of the
low-lying eigenmodes of the Dirac operator, from extended over the
whole system below the crossover, to localized on the scale of the
inverse temperature above the crossover~\cite{Garcia-Garcia:2006vlk,
  Kovacs:2012zq,Giordano:2013taa,Dick:2015twa,Ujfalusi:2015nha,
  Cossu:2016scb,Holicki:2018sms,Kehr:2023wrs,Meng:2023nxf} (see
Ref.~\cite{Giordano:2021qav} for a review). At high temperatures, a
``mobility edge'' separates localized low modes and delocalized bulk
modes; this decreases as a function of temperature, eventually
vanishing at a temperature in the crossover range. A similar behavior
has been observed in other gauge theories, both pure
gauge~\cite{Gockeler:2001hr,Gattringer:2001ia,Gavai:2008xe,
  Kovacs:2009zj,Kovacs:2010wx,Bruckmann:2011cc,Kovacs:2017uiz,
  Giordano:2019pvc,Vig:2020pgq,Bonati:2020lal,Alexandru:2021pap,
  Alexandru:2021xoi,Baranka:2021san,Baranka:2022dib,Alexandru:2023xho}
and in the presence of dynamical fermions~\cite{Giordano:2016nuu,
  Cardinali:2021fpu} and scalars~\cite{Baranka:2023ani}, with the
mobility edge vanishing at the critical point when the low- and
high-temperature regimes are separated by a genuine phase
transition~\cite{Kovacs:2017uiz,Giordano:2019pvc,Vig:2020pgq,
  Bonati:2020lal,Baranka:2021san,Baranka:2022dib,Giordano:2016nuu,
  Cardinali:2021fpu}.

While the Dirac eigenmodes obviously encode the whole dynamics of
quarks and antiquarks interacting with the non-Abelian gluon fields,
their relation with physical observables is often not straightforward,
as observables are generally obtained by integrating suitable
eigenvector correlators over the whole spectrum. A notable exception
is the density of near-zero modes, that in the chiral limit entirely
determines whether or not a chiral condensate is
developed~\cite{Banks:1979yr}, and even at physical quark masses is
mainly responsible for the fate of chiral symmetry. On the other hand,
it has been argued that the change in the localization properties of
near-zero modes is mostly due to the ordering of Polyakov loops above
the transition~\cite{Bruckmann:2011cc,Giordano:2015vla,
  Giordano:2016cjs,Giordano:2016vhx,Giordano:2021qav,
  Baranka:2022dib}, hinting at a close relation between localization
and the confining properties of the theory. Understanding the behavior
of the low Dirac modes across the transition is then key to unveiling
the connection between chiral symmetry restoration and deconfinement.

A thorough investigation of the ``geometric'' transition associated
with the appearance of localized low modes and of the corresponding
mobility edge in the Dirac spectrum, and of its relation with the
thermodynamic crossover, cannot dispense with the extrapolation to the
physical, continuum limit. Although so far no dedicated study of the
continuum scaling of localization properties has been performed in the
literature, there are both theoretical
arguments~\cite{Kovacs:2012zq,Giordano:2022ghy} and numerical evidence
that low-mode localization is not a lattice artifact, and survives the
continuum limit. In general, localization in QCD has been found with a
variety of discretizations of the Dirac operator (including
staggered~\cite{Garcia-Garcia:2006vlk,Kovacs:2012zq,Giordano:2013taa,
  Ujfalusi:2015nha}, M\"obius domain wall~\cite{Cossu:2016scb},
overlap on twisted mass~\cite{Holicki:2018sms,Kehr:2023wrs} and on
clover~\cite{Meng:2023nxf}). Moreover, the mobility edge in units of
the quark mass is a renormalization-group invariant
quantity~\cite{Kovacs:2012zq,Giordano:2022ghy}, that numerically shows
little dependence on the lattice
spacing~\cite{Kovacs:2012zq,Cardinali:2021fpu}.  Such numerical
evidence, however, relies on the connection between the localization
properties of eigenvectors and the statistical properties of the
corresponding eigenvalues~\cite{altshuler1986repulsion}, and was
obtained using the staggered discretization of the Dirac
operator. This combination is potentially problematic.

On the one hand, a lattice Dirac operator in the background of the
fluctuating gauge fields appearing in the path-integral formulation of
gauge theories is formally a (sparse) random matrix. As such, its
eigenvalues are expected to display the universal types of
correlations appearing in these systems~\cite{mehta2004random,
  Verbaarschot:2000dy}.  This allows one to identify the mobility edge
as the point in the spectrum where the spectral statistics 
switch
between the universal type corresponding to localized modes, and the
universal type corresponding to delocalized modes (which depends only
on the symmetry class of the lattice Dirac operator according to the
random matrix theory classification~\cite{Verbaarschot:2000dy,
  Zirnbauer:1996zz}). On the other hand, in four space-time dimensions
the staggered operator describes four ``tastes'' of fermions that
become exactly degenerate in the continuum limit, which in turn leads
the staggered spectrum to develop degenerate quartets of eigenvalues
as the lattice spacing decreases toward zero. The obvious additional
correlations between the nearly-degenerate eigenvalues appearing on
fine lattices then spoil the expected universal behavior. These
deviations are difficult to control theoretically, and make it more
difficult to reliably determine the localization properties of
eigenmodes using spectral statistics on finer lattices. This issue has
been known for a long time~\cite{Halasz:1996sb,Kovacs:2011km,
  Kovacs:2012zq},\footnote{For certain gauge groups and choices of
  representation for the fermions, the symmetries of the continuum
  Dirac operator differ from those of the staggered
  one~\cite{Verbaarschot:2000dy}: In these cases, the deviation from
  the expected universal behavior is actually needed, in order for the
  statistical properties of the lattice spectrum to approach those of
  the continuum spectrum (once the degeneracy is removed). However,
  deviations from the universal behavior will appear even when the
  staggered and continuum operator are in the same symmetry class.}
but it has never been fully addressed in the lattice literature so
far. A careful lattice study would be extremely interesting to assess
possible systematics affecting the calculation of the mobility edge.

In this respect, an important point to consider is that the influence
of the approximate taste degeneracy on the spectrum is reduced as the
volume is increased at fixed lattice spacing. Indeed, while the
typical distance between would-be-degenerate modes within the same
multiplet is controlled by the lattice spacing, the expected typical
distance between neighboring eigenvalues is controlled by the inverse
of the spectral density (i.e., the number of Dirac modes per unit
interval in the spectrum), and so by the inverse volume.  When
multiplets are clearly distinguishable, this quantity actually
measures the distance between them (up to the degeneracy factor).  As
soon as the lattice size becomes large enough and this distance
becomes comparable with the multiplet splitting, the multiplets
overlap and the corresponding structure gets washed away from the
spectrum. However, taste symmetry should still affect the
short-distance correlations between neighboring modes, as these still
carry a remnant of the multiplet structure.  Finally, when the inverse
spectral density becomes much smaller than the splitting within
would-be multiplets, taste symmetry effects disappear from
short-distance correlations; the inverse spectral density measures now
the distance between neighboring eigenvalues that, loosely speaking,
belong to different multiplets.  For the same reason, taste degeneracy
effects are less prominent where the spectral density is larger, such
as in the bulk of the spectrum, as well as near the origin at
temperatures below the pseudocritical temperature. On the other hand,
the low end of the staggered Dirac spectrum at high temperatures shows
a low spectral density, and taste-degeneracy effects on spectral
correlations are strong.\footnote{There is
  evidence~\cite{Kovacs:2017uiz,Kaczmarek:2023bxb,peak_wip_CB} that
  also the staggered spectrum develops a near-zero peak of eigenvalues
  on sufficiently fine lattices, similarly to what is observed using
  the overlap operator in the valence~\cite{Edwards:1999zm,
    Dick:2015twa,Alexandru:2015fxa,Alexandru:2019gdm,Alexandru:2019gdm,
    Vig:2021oyt,Kaczmarek:2021ser,Alexandru:2021pap,Alexandru:2021xoi,
    Meng:2023nxf}. In the latter case, near-zero peak modes also show
  peculiar localization properties~\cite{Alexandru:2021pap,
    Alexandru:2021xoi,Meng:2023nxf,Alexandru:2023xho}. The existence
  of such a peak in the Dirac spectrum in the presence of dynamical
  fermions is also supported by the arguments and the model
  calculations of Ref.~\cite{Kovacs:2023vzi}. However, the mobility
  edge discussed so far in the literature and in this paper is found
  in a spectral region far above this peak, beyond the depleted
  region.}

The discussion above shows that if we take the thermodynamic limit
before the continuum one, we can expect the distortions of the
spectral statistics due to taste degeneracy to disappear, and the
expected universal behavior based on the exact symmetries of the
staggered operator should emerge. This would in turn allow one to
employ standard methods to determine the localization properties of
the eigenmodes using spectral statistics, that can then be
extrapolated to the continuum. On the other hand, since the residual
fermion doubling of \textit{sea} staggered quarks is usually dealt
with using the so-called ``rooting trick''~\cite{Hamber:1983kx,
  Fucito:1984nu, Gottlieb:1988gr}, one expects the correct order of
limits to be continuum first and thermodynamic afterward. In this way, one
recovers an exact taste degeneracy in the sea lattice action, so that
taking the fourth root of the determinant in the path integral is
justified. However, as explained above, the uncontrolled effects of
taste degeneracy on the spectral statistics of the \textit{valence}
staggered operator make the determination of localization properties
and of the position of the mobility edge using these quantities
unreliable on finer lattices if the volume is not large enough. In
this context, a non-trivial question is whether one can actually
exchange the order of limits and still obtain the correct
results. It is important to note that exchanging limits
could be allowed when determining the mobility edge even though it is
not when studying spectral statistics. In fact, the mobility edge 
depends only on the localization properties of the eigenmodes, 
which are well defined entirely independently of the 
statistical properties of the spectrum. While localization properties 
and spectral statistics are related, it is the former that strongly 
affect the latter, and not the other way around.

Clearly, one could entirely bypass the problem by turning to the
direct study of the eigenvectors themselves. This is, however,
numerically more demanding, as it requires the use of several lattice
volumes, larger statistics, and a more sophisticated
finite-size-scaling analysis (see Refs.~\cite{PhysRevLett.105.046403,
  PhysRevB.84.134209,Ujfalusi:2015mxa}) to achieve the same
accuracy. Being able to reliably exploit the statistical properties of
the spectrum is then desirable, and this requires the exchange of
limits to be possible. In order for this to be also numerically
efficient, one needs that the dependence of the mobility edge on the
lattice spacing be mild, so that relatively coarse lattices, where the
distortion due to taste degeneracy is negligible, would suffice for a
continuum extrapolation. If this is the case, the results of
Refs.~\cite{Kovacs:2012zq,Cardinali:2021fpu}, that were obtained for
sufficiently large aspect ratios so as to avoid contamination from
eigenvalue multiplets, but at the same time on numerically manageable
volumes, could be fully trusted.

In this paper we study in detail the effects of taste degeneracy on
the statistical properties of the staggered Dirac spectrum, and how
they affect the determination of the mobility edge using spectral
statistics, with the ultimate goal of providing a controlled
investigation of the behavior of the mobility edge in the continuum
limit. After briefly reviewing the properties of the staggered
operator, and discussing localization and how to detect it using both
eigenvector properties and the statistical properties of the spectrum,
in Sec.~\ref{sec:locstag} we discuss the complications due to the
formation of nearly degenerate eigenvalue multiplets, and how the
lattice spacing and the lattice volume affect them. We then
argue that the continuum and thermodynamic limits can be exchanged when
studying the mobility edge. In Sec.~\ref{sec:numres} we present
numerical determinations of the mobility edge in the staggered
spectrum of high-temperature QCD for various lattice spacings and
volumes, obtained by a standard analysis of spectral statistics, and
study the dependence on the lattice spacing and on the lattice volume.
We then compare the results with a determination based on a direct
study of eigenvector properties, unaffected by taste degeneracy,
and with a heuristic estimate from the spectral statistics
of a suitably ``taste-symmetrized'' spectrum, where the approximate
taste degeneracy is removed by hand. Finally, we discuss the
extrapolation to the continuum limit.  We draw our conclusions in
Sec.~\ref{sec:concl}.

\section{Localization properties of staggered eigenmodes}
\label{sec:locstag}

This section is devoted to briefly summarizing the properties of the
staggered discretization of the Dirac operator~\cite{Kogut:1974ag,
  Susskind:1976jm,Banks:1976ia}, as well as the main techniques and
quantities used to study the localization properties of its
eigenmodes. We then discuss in some detail the issues related to taste
degeneracy.

\subsection{The staggered lattice Dirac operator} 
\label{sec:stag_def}

The massless staggered operator in lattice QCD reads
\begin{equation}
  \label{eq:stag_op}
  a  D^{\mathrm{stag}}[U] = \f{1}{2}\sum_{\mu=1}^4 \eta_\mu \left(U_\mu
    \mathrm{T}_\mu -     \mathrm{T}_\mu^\dag U_\mu^\dag\right)\,,
\end{equation}
where $U_\mu(x)\in\mathrm{SU}(3)$, with $x=(x_1,x_2,x_3,x_4)$
and $\mu=1,\ldots,4$, is a gauge link variable attached to the link
connecting the sites $x$ and $x+a\hat{\mu}$ of a
$N_s\times N_s\times N_s\times N_t$ hypercubic lattice with lattice
spacing $a$; $\mathrm{T}_\mu$ are unit translation operators with
periodic boundary conditions in the spatial directions $\mu=1,2,3$
and antiperiodic boundary conditions in the temporal direction
$\mu=4$; and $\eta_\mu$ are the staggered phases,
$\eta_\mu(x)=(-1)^{\sum_{\alpha<\mu} x_\alpha}$. The operator
$D^{\mathrm{stag}}$ is anti-Hermitean, so its eigenmodes obey
\begin{equation}
  \label{eq:stag_op2}
  D^{\mathrm{stag}}[U]\psi_n[U] = i\lambda_n[U]\psi_n[U]\,, 
\end{equation}
with $\lambda_n\in\mathbb{R}$. For notational simplicity we will
generally drop the dependence on the gauge configuration. These
eigenmodes carry a spacetime index $x$, running over the lattice
sites, and a colour index $c=1,2,3$. Thanks to the chiral property
$\{\varepsilon,D^{\mathrm{stag}}\}=0$, where
$\varepsilon(x)=(-1)^{\sum_{\alpha} x_\alpha}$, the spectrum of
$D^{\mathrm{stag}}$ is symmetric about zero, with
$\psi_{-n}\equiv \varepsilon\psi_n$ obeying
$D^{\mathrm{stag}}\psi_{-n}=-i\lambda_n\psi_{-n}$.

The operator $D^{\mathrm{stag}}$ is formally $i$ times a random
Hamiltonian, like those used in condensed matter physics to model
systems with disorder (see, e.g., Refs.~\cite{lee1985disordered,
  kramer1993localization}). Here the disorder is provided by the
fluctuations of the gauge fields, over which one integrates in the
lattice formulation of gauge theories. The probability distribution of
the entries of our random Hamiltonian is then determined by the
specifics of the discretization of the Yang-Mills action for the gauge
fields, as well as by those of the improvement techniques employed to
speed up the approach to the continuum. In particular, the link
variables $U_\mu(x)$ corresponding to the discretized non-Abelian
gauge fields need not be (and are not in current numerical practices)
the same ones appearing in the lattice Yang--Mills action. These
details are not relevant to the general discussion, and are presented
below in Sec.~\ref{sec:numressetup}. For our purposes it suffices to
say that the integration over gauge fields defines an expectation
value $\la\ldots\ra$, corresponding to the disorder average in the
language of disordered systems, with the usual properties
$\la 1 \ra =1$ and $\la \Oc[U] \Oc[U]^*\ra \ge 0$ for a generic
observable $\Oc$ that depends only on the gauge fields.

\subsection{(Inverse) participation ratio and fractal dimension}
\label{sec:IPR}

The localization properties of the eigenmodes of the staggered Dirac
operator are studied in full analogy with those of the eigenmodes of
random Hamiltonians~\cite{lee1985disordered,kramer1993localization}.
In qualitative terms, the localization properties of the eigenmodes of
such systems are defined by the scaling of their effective size,
averaged over disorder realizations, with the size of the system. For
localized modes, the average mode size remains constant in the large
volume limit, while for delocalized modes it grows proportionally to
the volume.

To make these statements quantitative, one introduces the inverse
participation ratio (IPR) of the eigenmodes, defined here in a
gauge-invariant way as
\beq
\label{eq:stag_op3}
\IPR_n &\equiv& \sum_x \Vert \psi_n(x)\Vert^4, \\
\Vert \psi_n(x)\Vert^2 &\equiv& \sum_{c=1}^{3} \vert
\psi^c_n(x)\vert^2\,,
\eeq
where the latter sum is performed over color indices. It is understood
that modes are normalized, $\sum_x \Vert \psi_n(x)\Vert^2 = 1$. The
effective mode size is simply $\IPR^{-1}$: indeed, for a mode with
constant amplitude square $\Vert \psi(x)\Vert^2=1/V_0$ in a region of
size $V_0$ (in lattice units), one finds $\IPR^{-1}=V_0$, while for a
fully delocalized mode with $\Vert \psi(x)\Vert^2=1/(V_sN_t)$, where
$V_s=N_s^3$ is the spatial volume in lattice units, one finds
$\IPR^{-1}=V_sN_t$. Instead of the mode size, it is sometimes
convenient to use the participation ratio (PR), i.e., the fraction of
lattice volume effectively occupied by the mode:
\beq
\label{eq:stag_op4}
\PR_n \equiv \f{\IPR_n^{-1}}{V_s N_t}\,.
\eeq
Notice that $\psi_{-n}$ and $\psi_n$ have the same IPR, so the
localization properties are exactly the same for the positive and
negative part of the spectrum.

The localization properties of eigenmodes in a given spectral region
are determined by the scaling of the PR in the thermodynamic limit,
after averaging over gauge configurations (i.e., over disorder
realizations),
\beq
\label{eq:stag_op5}
\overline{\PR}(\lambda,N_s)\equiv\f{1}{\rho_0(\lambda)}\left\la
\sum_n\delta(\lambda-\lambda_n) 
\PR_n\right\ra\,.
\eeq
Here
\beq
\label{eq:un_specdens}
\rho_0(\lambda) \equiv  \left\la\sum_n\delta(\lambda-\lambda_n)\right\ra
\eeq
is the (non-normalized) spectral density, which is expected to scale
proportionally to $V_s N_t$ in the large-volume limit. In
Eq.~\eqref{eq:stag_op5} we have made the dependence of
$\overline{\PR}$ on $N_s$ explicit, while leaving out that on $N_t$,
as the thermodynamic limit is taken here at fixed $N_t$. In spectral
regions where modes are localized, $\overline{\PR}$ tends to zero as
$1/V_s$ as the spatial volume is increased, while for delocalized
modes it tends to a constant. The large-volume behavior of
$\overline{\PR}$ determines the fractal dimension $\alpha$ of the
modes ($D_2$ in the notation of
Refs.~\cite{Evers:2008zz,Ujfalusi:2015mxa}), defined locally in the
spectrum as
\beq
\label{eq:stag_op6}
\alpha(\lambda) = 3+ \lim_{N_s\to\infty} \f{\log
\overline{\mathrm{PR}}(\lambda,N_s)}{\log N_s}\,,
\eeq
with $\alpha=0$ for localized modes and $\alpha=3$ for delocalized
modes. Any intermediate behavior corresponds to ``critical modes'' in
the language of condensed matter physics~\cite{Evers:2008zz}.

The discussion above deals with the lattice system at fixed $N_t$. To
study localization in the continuum limit at fixed temperature, one
can either proceed as above, taking $N_s\to\infty$ at fixed $a$ and
$T=1/(aN_t)$; or take first $a\to 0$ at fixed $L=aN_s$ and $T$ to
define a continuum system, and only after that study the scaling of
the PR with $L$.

\subsection{Spectral statistics}
\label{sec:specstat}

The localization properties of the eigenmodes of a random Hamiltonian
determine the statistical properties of the corresponding
eigenvalues~\cite{altshuler1986repulsion}. To see this, consider a
specific disorder realization and the corresponding basis of
eigenvectors, and perturb the disorder configuration by adding some
local fluctuation. Once represented in the basis of the unperturbed
eigenvectors, the perturbed Hamiltonian will have nonzero off-diagonal
matrix elements for any pair of delocalized modes, while off-diagonal
matrix elements involving localized modes can be non-negligibly
different from zero only if they are localized where the fluctuation
is introduced. In other words, delocalized modes are easily mixed by
disorder fluctuations, with a dense matrix describing this mixing,
while localized modes are sensitive only to disorder fluctuations
appearing near their localization region. The statistical properties
of eigenvalues associated with delocalized modes are then expected to
match those of the appropriate Gaussian ensemble of Random Matrix
Theory (RMT), according to the symmetry class of the disordered
Hamiltonian~\cite{mehta2004random,Verbaarschot:2000dy}. For
uncorrelated disorder, or for disorder with only short-range
correlations, localized modes are instead expected to fluctuate
independently and thus to obey Poisson statistics.

A convenient observable in this context is the probability
distribution of the so-called unfolded level spacings, i.e., the
distance between neighboring eigenvalues measured on the scale of the
average distance between neighboring eigenvalues found in the relevant
spectral region~\cite{mehta2004random,Verbaarschot:2000dy}. Formally,
one defines the unfolded spectrum via the mapping
\beq
\label{eq:unfold1}
\lambda_n \rightarrow x_n = \int_{\lambda_{\min}}^{\lambda_n}
d\lambda\, \rho_0(\lambda)\,, \eeq
which, by construction, has unit spectral density anywhere, and the
corresponding unfolded spacings as $ s_n \equiv x_{n+1}-x_n$. When the
volume becomes large, the probability distribution $p(s;\lambda,N_s)$
of the unfolded spacings, computed locally in the spectrum,
\beq
\label{eq:unfold2}
p(s;\lambda,N_s) \equiv
\f{1}{\rho_0(\lambda)}\left\la
\sum_n\delta(\lambda-\lambda_n) 
\delta(s-s_n)
\right\ra\,,
\eeq
is expected in general to depend only on the localization properties
of the modes in the spectral region of interest, and on the symmetry
class of the corresponding random Hamiltonian. For localized modes and
Poisson statistics one expects an exponential distribution,
\beq
\label{eq:unfold3}
p_{\mathrm{P}}(s) = e^{-s}\,,
\eeq
while for delocalized modes one expects the distribution to be the
same as the one found in the appropriate Gaussian ensemble of RMT.
These are known but not available in closed form, and often
approximated by the so-called ``Wigner surmise'',
\beq
\label{eq:unfold4}
p_{\mathrm{RMT}}(s)\simeq p_{\mathrm{WS}}(s) = a_\beta s^\beta
e^{-b_\beta s^2}\,, \eeq
where $\beta=1,2,4$ for the orthogonal, unitary, and symplectic
symmetry class, respectively. The coefficients $a_\beta,b_\beta$ are
determined by the normalization conditions $\int ds\,p(s)=1$ and
$\la s\ra =\int ds\,p(s)s=1$, the latter following from the fact that
the unfolded spectrum has unit spectral density. In the case of
lattice QCD with staggered quarks, transforming under the fundamental
representation of the gauge group $\SU(3)$, the relevant symmetry
class is the unitary class.\footnote{On general grounds, for
  $\SU(N_c)$ lattice gauge theories with fermions transforming
  according to the fundamental representation of the gauge group, the
  staggered operator is in the unitary class if $N_c\ge 3$, and in the
  symplectic class if $N_c=2$. For $N_c=2$ this differs from the
  symmetry class of the continuum Dirac operator, which is the
  orthogonal one~\cite{Verbaarschot:2000dy}.}

An exception to this classification is represented by the mobility
edge separating localized and delocalized modes. At the mobility edge
the localization length diverges, and the system undergoes a
second-order phase transition in the spectrum, known as Anderson
transition~\cite{Evers:2008zz}. Correspondingly, eigenmodes become
critical, with a characteristic fractal dimension $\alpha_*$ that
depends on the symmetry class~\cite{Ujfalusi:2015mxa}, and the
corresponding eigenvalues obey a characteristic type of statistics
$p_*(s)$, intermediate between Poisson and
RMT-type~\cite{al1988repulsion,Shklovskii:1993zz,Evers:2008zz}.

The discussion above is valid only in the limit of infinite volume,
where any overlap between different localized modes becomes entirely
negligible, and where delocalized modes can fully spread out on a
system of infinite size. In any finite volume there are instead
corrections to the Poisson or RMT-type behavior, that tend to zero as
the volume is increased. In the presence of a mobility edge one then
expects $p(s;\lambda,N_s)$ to interpolate between a near-exponential
behavior and a near-RMT behavior, passing through the critical
behavior at the mobility edge, where the properties of the eigenmodes
are expected to be scale-invariant. Then, as $N_s$ is increased,
$p\to p_{\mathrm{P}}$ in the localized spectral region,
$p\to p_{\mathrm{RMT}}$ in the delocalized spectral region, and
$p=p_*$ at the mobility edge $\lambda=\lambda_c$.  This can be used to
determine $\lambda_c$ and the critical exponents of the Anderson
transition by means of a finite size scaling
analysis~\cite{Shklovskii:1993zz}, as was done in
Ref.~\cite{Giordano:2013taa} for QCD.  Alternatively, if the critical
distribution $p_*$ is known (as is the case, to some extent, for the
unitary class~\cite{Nishigaki:2013uya}), one can use it to determine
$\lambda_c$ as the point where the spectral statistics matches the
critical one. This can be done very efficiently using a single lattice
volume: by virtue of the scale-invariance of the critical point, such
an estimate is expected to suffer only from very little finite-size
effects.

As a concrete example, one can monitor how the integrated distribution
$I_{s_0}$,
\beq
\label{eq:unfold5}
I_{s_0}(\lambda,N_s) = \int_0^{s_0} ds\,p(s;\lambda,N_s)\,,
\eeq
for a suitably chosen $s_0$, varies along the spectrum, and where it
matches the critical value. For the unitary class one customarily
chooses $s_0\simeq 0.508$ to maximize the difference between the
expected values for Poisson and RMT-type statistics, i.e.,
$I_{s_0,\mathrm{P}}\simeq 0.398$ and
$I_{s_0,\mathrm{RMT}}\simeq 0.117$; for this choice of $s_0$ the
critical value $I_{s_0,c}$ has been determined in
Ref.~\cite{Giordano:2013taa}.  One could similarly use the second
moment of the distribution, $\la s^2\ra $,
\begin{equation}
\label{eq:unfold6}
\la s^2\ra (\lambda,N_s)
= \int_0^\infty ds\,s^2\,p(s;\lambda,N_s)\,,
\end{equation}
or equivalently the variance
$\Var(s) \equiv \la s^2\ra - \la s\ra^2 $, or any
other feature extracted from the unfolded level spacing
distribution. For $\Var(s)$, the expectations for Poisson and for RMT
statistics in the unitary class are $\Var(s)_{\mathrm{P}}=1$ and
$\Var(s)_{\mathrm{RMT}}=\f{3\pi}{8}-1$, respectively.

The mobility edge can then be estimated as the point
$\lambda_c^{\mathrm{stat}}$ where $I_{s_0}$, or any other observable,
intercepts the critical value, i.e.,
$I_{s_0}(\lambda_c^{\mathrm{stat}},N_s)=I_{s_0,c}$. In general,
$\lambda_c^{\mathrm{stat}}$ depends on the spatial size of the system,
$N_s$, although this dependence is expected to be mild. In the absence
of approximate symmetries, this is expected to be the case also for
the lattice Dirac spectrum at fixed finite temperature
$T=1/(aN_t)$. In this case $\lambda_c^{\mathrm{stat}}$ depends on $T$
as well as on the lattice spacing, $a$, or, equivalently, on the
temporal extension, $N_t$. For the renormalization-group invariant
combination
$\lambda_c^{\mathrm{stat}}/m_s$~\cite{Kovacs:2012zq,Giordano:2022ghy}
also the dependence on $N_t$ is expected to be mild. We discuss now
how these general expectations are modified in the case of staggered
fermions.

\subsection{Effects of taste degeneracy}
\label{sec:spec_taste}

The discussion above ignored entirely the possible presence of
additional correlations between eigenvalues due to peculiarities in
the structure of the random Hamiltonian. Such correlations can be
induced by the presence of an approximate symmetry, leading to
near-degenerate multiplets of eigenvalues. In this case, the
corresponding symmetry-breaking parameter controls the typical
splitting $\delta$ within the would-be-degenerate multiplets, which,
in turn, gives the spectral scale at which their effects are felt.
When this scale is smaller than or comparable to the expected distance
$\Delta=1/\rho_0$ between neighboring eigenvalues, it affects the
unfolded level spacing distribution, skewing it toward lower values:
Indeed, neighboring levels belonging to the same approximate
multiplet prefer to stay closer than what one would expect simply
based on the inverse of the spectral density, as their distance is
controlled by a different mechanism with a corresponding smaller
spectral scale.  More precisely, one has clearly separated multiplets
of $n$ eigenvalues if their size $(n-1)\delta$ is much smaller than
the typical distance $n/\rho_0 = n\Delta$ between them, i.e., if
$\delta\ll \Delta$. In this case, out of $n$ level spacings, $n-1$ are
of order $\delta$, and one is of order $n\Delta-(n-1)\delta$.  This
means that the average level spacing results from the averaging of
fluctuations around two different typical values; for this reason, one
expects also a larger variance for the unfolded level spacing
distribution.

On the other hand, the spectral density is proportional to the volume,
and so in the thermodynamic limit at fixed symmetry-breaking parameter
many approximate multiplets will overlap. Indeed, when
$\delta \sim \Delta$, the size of a multiplet becomes comparable with
the distance between multiplets, and their clear separation becomes
impossible. Nonetheless, at this stage the eigenvectors are likely to
still carry a remnant of the multiplet structure, that reflects on the
correlations between the corresponding eigenvalues, including
neighboring ones. When $\delta \gg \Delta$, many multiplets overlap:
the approximate symmetry will still affect the eigenvalue
correlations, but the corresponding effects will be visible only on a
spectral scale much larger than the typical separation between
neighboring eigenvalues. Loosely speaking, assuming that an assignment
of eigenvalues to multiplets still makes sense in such a dense
environment, one finds that neighboring eigenvalues almost certainly
belong to different multiplets. This means that short-range
correlations involving eigenvalues at a fixed separation in mode
number, such as the ones governing the unfolded level spacing
distribution, are entirely unaffected by the approximate symmetry in
the thermodynamic limit. Loosely speaking again, one would be
measuring the distribution of the level spacing between members of
different multiplets, for which there should be no distortion from the
universal expectation.

The staggered operator provides precisely an example of the situation
discussed above: while in the continuum limit it has an exact symmetry
under the exchange of degenerate tastes, on a finite lattice this
symmetry is only approximate. This manifests in the formation of
nearly-degenerate eigenvalue multiplets as one gets sufficiently close
to the continuum. The symmetry-breaking parameter is here the lattice
spacing $a$, and the splittings $\delta$ in the nearly-degenerate
eigenvalue multiplets are of order $\delta \sim (a\Lambda)^2\Lambda$
(see, e.g., Ref.~\cite{Lepage:2011vr,Donald:2011if}), with $\Lambda$
some physical mass scale. In the thermodynamic limit at fixed $a$, the
argument above applies: the multiplet structure gets washed out, and
its effects on short-range eigenvalue correlations disappear. On the
other hand, as $a\to 0$ the splittings decrease and multiplets become
more degenerate, and more distinguishable from each other: this effect
competes with the wash-out effect of the thermodynamic limit, and so
the order in which the continuum and the thermodynamic limits are
taken becomes important.

In finite-temperature calculations, it is customary to take the
continuum limit $a\to 0$ at fixed temperature $T=1/(aN_t)$ and fixed
aspect ratio $LT=N_s/N_t$, corresponding to keeping the lattice
spatial volume (as well as its temporal extension) fixed in physical
units. In this case, the relative size of the multiplet splittings
$\delta$ and the typical level spacing
$\Delta = 1/\rho_0 \sim (T/L^3)/\Lambda^3$ is
\begin{equation}
  \label{eq:dDest}
\begin{aligned}
  \f{\delta}{\Delta} & \sim (a\Lambda)^2
  \left(LT\right)^3\left(\f{\Lambda}{T}\right)^4\,,
\end{aligned}
\end{equation}
i.e., it is expected to decrease quadratically with $a$. This means
that short-range correlations become more sensitive to the effects of
taste degeneracy as $a\to 0$, leading to significant systematic
effects in the study of spectral statistics, and sizeable deviations
from the expected universal behavior of the unfolded spectrum, as soon
as $\delta\lesssim \Delta$. 

The distortion of the unfolded level spacing distribution caused by
the emergence of nearly-degenerate eigenvalue multiplets leads to
values of $I_{s_0}$ and $\mathrm{Var}(s)$ larger than one would expect
based on the localization properties of the eigenmodes, and an
estimate $\lambda_c^{\mathrm{stat}}$ of the mobility edge
using the intercept with the critical value of some spectral
statistic, as explained above, would generally lead to an
overestimation of the actual value $\lambda_c^{\text{true}}$. The
actual physical value of the mobility edge is, of course, independent
of the use of spectral statistics to determine it, as it originates in
the properties of the eigenvectors.  In this context, its most
important aspect is its characterization as the point in the spectrum
where the localization length diverges and a second-order Anderson
transition takes place.  This allows one to determine it, for example,
by using the intercept of the local fractal dimension
$\alpha(\lambda)$ with the corresponding critical value $\alpha_*$.

For this reason, the determination of $\lambda_c$ from spectral
statistics is expected to converge to the true position of the
mobility edge if one takes the thermodynamic limit at fixed lattice
spacing. In fact, as explained above, by taking limits in this order
one gets rid of the taste-multiplet problem, and the spectral
statistics are expected to behave like those of an ordinary system
without approximate symmetries. Then, the scale-invariant nature of
the physics of critical Dirac eigenmodes should reflect in the
statistical properties of the spectrum in the usual way. This means
that the spectral statistics at the mobility edge should be
independent of the volume for sufficiently large lattice sizes (as
soon as distortions due to taste symmetries become negligible), and
agree with the universal critical statistics expected for the given
symmetry class (irrespectively of approximate symmetries of the
system). Conversely, a scale-invariant behavior of the spectral
statistics is a sign of a second-order transition in the spectrum, and
so of the presence of a \textit{bona fide} mobility edge.  In other
words, one expects that at fixed $N_t$,
$\lambda_c^{\mathrm{stat}}=\lambda_c^{\mathrm{stat}}(T;N_t,N_s)\to
\lambda_c^{\text{true}}(T;N_t)$ as $N_s\to \infty$; and that if $N_s$
is large enough then $\lambda_c^{\mathrm{stat}}(T;N_t,N_s)$ shows a
plateau in $N_s$, entirely unaffected by taste-degeneracy
effects. This, of course, should be verified explicitly.

However, even if this is the case, this approach rapidly gets
numerically very intensive both for the generation of gauge
configurations and for the diagonalization of the staggered operator
as one gets closer to the continuum, until it becomes impossible to
reach sufficiently large aspect ratios to see the plateau, and the
extrapolation of the results for the mobility edge toward the
thermodynamic limit introduces larger systematic effects. As the
lattice spacing is reduced the statement that the approach based on
spectral statistics is numerically more efficient becomes
questionable.

Clearly, the ideal scenario would be that indeed
$\lambda_c^{\mathrm{stat}}(T;N_t,N_s)$ tends to the true mobility edge
as the volume is increased; that the continuum and thermodynamic
limits commute; and that the true mobility edge in units of the quark
mass, $\lambda_c^{\text{true}}(T;N_t)/m_s$, depends only mildly on the
lattice spacing,
$\lambda_c^{\text{true}}(T;N_t)/m_s\simeq
\lambda_c^{\text{cont}}(T)/m_s^{\text{cont}} =
\lim_{N_t\to\infty}\lambda_c^{\text{true}}(T;N_t)/m_s $, so that
relatively coarse lattices, for which distortion effects due to taste
degeneracy are negligible, can be used for a continuum extrapolation.

\subsection{Continuum and thermodynamic limits}
\label{sec:spec_taste_cont}

While the discussion above shows that continuum and thermodynamic
limits do not commute for spectral statistics (unless possibly if one
takes care of the approximate taste degeneracy), it does not exclude
this possibility for the mobility edge. In fact, the mobility edge is
directly related to the localization properties of the Dirac
eigenmodes, and only indirectly to spectral statistics, through the
influence that localization properties have on them; one could
identify it by looking only at eigenvector properties, without any
reference to spectral statistics. We now argue that the two limits can
indeed be exchanged when studying localization properties and the
position of the mobility edge.

If one switched off the taste-breaking part of the staggered operator,
one would find degenerate eigenspaces of ``unperturbed'' modes related
by taste symmetry transformations.  Since these act on the scale of an
elementary hypercube, mixing the corresponding components of the quark
wave function, the modes in a degenerate eigenspace will have similar
localization properties; in particular, all the modes in the
eigenspace corresponding to a localized mode will be localized in the
same region and will have similar sizes. In the symmetric theory, one
could then straightforwardly define an ``unperturbed'' mobility
edge. The question is how this changes when taste-breaking effects are
taken into account.

To answer this question, we set
$H = -iD^{\mathrm{stag}}/\Lambda = H_0 + g\Delta$, where $H_0$ is the
taste-symmetric part, $\Delta$ the taste-breaking part, $\Lambda$ a
typical QCD scale, and $g=(a\Lambda)^2$ plays the role of
dimensionless coupling.
In dimensionless units, we have $E_n=\lambda_n/\Lambda$ for the
eigenvalues of $H$, and $L_\Lambda=L\Lambda$ and
$V_\Lambda=(L\Lambda)^3=V\Lambda^3$ for the system size.
To find the
exact spectrum $\{E_n\}$ and eigenvectors $\{\psi_n\}$ of $H$, we work
in the basis $\{\psi_n^{(0)}\}$ of ``unperturbed'' eigenvectors of $H_0$,
\begin{equation}
  \label{eq:pt1}
  H \psi_n = E_n \psi_n\,, \qquad
  H_0 \psi_n^{(0)} = E_n^{(0)} \psi_n^{(0)}\,.
\end{equation}
In this basis we have to diagonalize the matrix
\begin{equation}
  \label{eq:pt2}
  \begin{aligned}
  H_{n'n} &= (\psi_{n'}^{(0)},H\psi_n^{(0)}) =
  E_n^{(0)}\delta_{n'n}  + g(\psi_{n'}^{(0)},\Delta\psi_n^{(0)}) \\ &=
  E_n^{(0)}\delta_{n'n}  + g\Delta_{n'n}\,,\\
  (\phi,\psi) &\equiv \sum_{x}\sum_{c=1}^3 \phi^c(x)^*\psi^c(x)\,.
  \end{aligned}  
\end{equation}
For definiteness, within degenerate subspaces we choose eigenvectors
that diagonalize $\Delta$ restricted to the subspace, i.e., $\Delta_{n'n} = 0$ if
$n'\neq n$ but $E_{n'}^{(0)}=E_n^{(0)}$.

We work with fine and large lattices, so that the scale of the 
taste-breaking interaction, $g$, and that of the inverse spectral
density, $1/V_\Lambda$, are both small, but we do not initially
specify which one is smaller. As discussed in the previous subsection,
these are the two scales relevant to the status of taste-symmetry
effects in the exact spectrum, and their relative size distinguishes
two regimes: if $g< 1/V_\Lambda$, quartets are well-formed, with
$O(g)$ intraquartet splitting and $O(1/V_\Lambda)$ distance between
quartets; if $1/V_\Lambda <g$, quartets are overlapping and level
spacings are generally $O(1/V_\Lambda)$. The level spacing
between non-degenerate unperturbed eigenvalues is, in any case, of
order $O(1/V_\Lambda)$.

The unperturbed eigenvectors $\{\psi_n^{(0)}\}$ are in part localized
and in part delocalized, but working in a finite volume the
distinction is, of course, ambiguous. One can nonetheless provide a
finite-volume estimate of the unperturbed mobility edge $E_c^{(0)}$
using, e.g., the fractal dimension, or even spectral statistics, that
are unproblematic here since one can remove the exact degeneracy by
hand.  Using standard scaling arguments (see, e.g.,
Ref.~\cite{Giordano:2013taa}), the difference between such estimates
and the infinite-volume value is expected to be not more than of the
order $O(1/L_\Lambda^{1/\nu})$, where $\nu$ is the correlation length
critical exponent of the Anderson
transition~\cite{lee1985disordered,kramer1993localization,Evers:2008zz},
and in practice much less than that. In fact, the localized or
delocalized nature of modes in a finite volume is ambiguous at most
within the scaling region
$|E^{(0)}-E_c^{(0)}|L_\Lambda^{1/\nu} = O(1)$ around the mobility
edge, and truly so only within the much smaller region where
corrections to scaling are important, and one cannot decide the nature
of the modes based on the leading behavior.

Switching the taste-breaking interactions back on, the unperturbed
eigenvectors will mix, and localization properties may change. To see
to what extent this is the case, we need to estimate the order of
magnitude of the matrix elements $\Delta_{n'n}$.
For $n$ and $n'$ both localized, $\Delta_{n'n}$ receives contributions
only from the spacetime region where the two localized modes
overlap.
Since localization is typically exponential
one expects 
\begin{equation}
  \label{eq:pt3_1}
  \Delta_{n'n} \sim
  \left(\f{\xi_n\xi_n'}{\xi_{n'n}^2}\right)^{\f{3}{2}}e^{-\f{d_{n'n}}{\xi_{n'n}}}\,, \qquad
  \xi_{n'n}^2 = \xi_n^2 + \xi_{n'}^2\,,
\end{equation}
with $\xi_{n},\xi_{n'}$ the localization lengths of the two modes,
and $d_{n'n}$ the distance between their localization centers.\footnote{One
expects different modes to be localized in the same region only if
this is demanded by symmetry reasons, so this is expected only for
modes in the same degenerate subspace. For these $\Delta_{n'n}$ has
been set to zero by a judicious choice of basis.}
The largest matrix elements correspond to pairs of modes whose
localization regions are close, but whose eigenvalues are typically
far away from each other in the spectrum. Conversely, eigenvalues that
are close in the spectrum correspond typically to modes that are far
away in space: since on average they are distributed uniformly in
space, the chance that they are nearby is $O(1/N_{\mathrm{loc}}) =
O(1/V_\Lambda)$, where
$N_{\mathrm{loc}}$ is the number of localized modes.

For $n$ localized and $n'$ delocalized, $\Delta_{n'n}$ receives
contributions only from the localization region of $\psi_n^{(0)}$,
and since $\psi_{n'}^{(0)} \sim 1/\sqrt{V}$,
one has $ \Delta_{n'n}\sim
    \sqrt{ {\xi_n^3}/{V} }$.
In a region of size $O(1/L_\Lambda^{1/\nu})$ around $E^{(0)}_c$,
where localized modes have size $\xi_n^3 = O(V)$, this can be as
large as $O(1)$; however, $H_{n'n}$ is further suppressed by a
factor of $g$.
Finally, for $n$ and $n'$ both delocalized,
$\Delta_{n'n}$ receives contributions from the whole volume, and is of
order $\Delta_{n'n}\sim V /(\sqrt{V})^2= 1$.

In the delocalized part of the unperturbed spectrum, the effect of the
taste-breaking interaction is expected to wash out the quartet
structure, despite the smallness of $g$, if $1/V_\Lambda < g$: in this
case the typical unperturbed level spacing is smaller than the effect
of the perturbation, and mixing affects members of different
quartets. The taste-breaking interaction is otherwise expected not to
change much neither the localization properties of the modes, nor to
shift much the spectrum. In fact, this interaction amounts in practice
to adding some weak off-diagonal disorder to the system, as the
perturbation $\Delta$ has only off-diagonal matrix elements of order 1
(as they are built out of the unitary gauge link variables), and
their order of magnitude does not fluctuate across the lattice.
This type of disorder is known to be largely ineffective in localizing
modes except in the tails of the spectrum, see
Refs.~\cite{ECONOMOU1977285,WEAIRE1977863,PhysRevB.74.113101}.  Here
it is the size of the fluctuations of $\Delta$ that is relevant, and not the
relative size of $g$ and $1/V_\Lambda$, so taste symmetry plays no
role. One then expects that if the mobility edge moves, 
it does not do so because of interference effects between strongly
mixing delocalized modes leading to localization, but 
because of mixing between localized and delocalized
modes, or between localized modes alone leading to delocalization.

Indeed, using the obvious identity
\begin{equation}
  \label{eq:pt4}
  (E_n -E_{n'}^{(0)})(\psi_{n'}^{(0)},\psi_n) = g(\psi_{n'}^{(0)},\Delta\psi_n)\,,
\end{equation}
and denoting by
$\Pi^{(0)}_{\mathrm{loc/deloc}}=\sum_{n'\in\mathrm{loc/deloc}}\psi_{n'}^{(0)}\psi_{n'}^{(0)\dag}$
the projectors on the localized or delocalized unperturbed
eigenvectors, 
one has for any $\psi_n$ the exact bounds
\begin{equation}
  \label{eq:pt5_alt2}
  \begin{aligned}
    & g^2 \sum_{n'}|(\psi_{n'}^{(0)},\Delta\psi_n)|^2 = \sum_{n'}(E_n
    -E_{n'}^{(0)})^2|(\psi_{n'}^{(0)},\psi_n)|^2 \\ &\ge
      \min_{n'\in\mathrm{loc}}(E_n -E_{n'}^{(0)})^2
      \sum_{n'\in\mathrm{loc}}|(\psi_{n'}^{(0)},\psi_n)|^2  \\
    &=
      \Vert(1-\Pi^{(0)}_{\mathrm{deloc}})\psi_n\Vert^2\min_{n'\in\mathrm{loc}}(E_n
      -E_{n'}^{(0)})^2 \\ &\ge
    \left(1-\Vert\Pi^{(0)}_{\mathrm{deloc}}\psi_n\Vert^2\right)\min_{n'\in\mathrm{loc}}(E_n
    -E_{n'}^{(0)})^2 \,,
  \end{aligned}
\end{equation}
where minimization is restricted to the localized unperturbed
spectrum, and
\begin{equation}
  \label{eq:pt5_alt}
  \begin{aligned}
    & g^2 \sum_{n'}|(\psi_{n'}^{(0)},\Delta\psi_n)|^2 = \sum_{n'}(E_n
    -E_{n'}^{(0)})^2|(\psi_{n'}^{(0)},\psi_n)|^2 \\ &\ge
      \min_{n'\in\mathrm{deloc}}(E_n -E_{n'}^{(0)})^2
      \sum_{n'\in\mathrm{deloc}}|(\psi_{n'}^{(0)},\psi_n)|^2  \\
    &=
      \Vert(1-\Pi^{(0)}_{\mathrm{loc}})\psi_n\Vert^2\min_{n'\in\mathrm{deloc}}(E_n
      -E_{n'}^{(0)})^2 \\ &\ge
    \left(1-\Vert\Pi^{(0)}_{\mathrm{loc}}\psi_n\Vert^2\right)\min_{n'\in\mathrm{deloc}}(E_n
    -E_{n'}^{(0)})^2 \,,
  \end{aligned}
\end{equation}
where minimization is restricted to the delocalized unperturbed
spectrum.

Due to the weak localizing power of the taste-breaking interaction,
one expects that if $\psi_n$ is a localized exact mode, then it does
not come entirely from the mixing of delocalized unperturbed modes, so
that $1-\Vert\Pi^{(0)}_{\mathrm{deloc}}\psi_n\Vert^2$ is a nonzero
number of order $O(1)$ (except possibly in the tails 
of the spectrum). One finds then from Eq.~\eqref{eq:pt5_alt2} that for a
localized exact mode, $E_n < E_c^{(0)} + \delta_+$, with $\delta_+$ a
positive number of order $O(g)$.

On the other hand, Eq.~\eqref{eq:pt5_alt} shows that if $\psi_n$ is a
delocalized exact mode, then it can be found below the unperturbed
mobility edge at distance of order $O(g^{1-\epsilon})$, $\epsilon\le 1$,
only if
$1-\Vert\Pi^{(0)}_{\mathrm{loc}}\psi_n\Vert^2=O(g^{2\epsilon})$, i.e.,
if a delocalized $\psi_n$ is obtained essentially only by forming a
linear combination of localized unperturbed modes, up to a
contribution from delocalized unperturbed modes of order
$O(g^{\epsilon})$, i.e.,
\begin{equation}
  \label{eq:pert_delovec}
  \psi_n = \Psi_{\mathrm{loc}}^{(0)} + g^{\epsilon}\Psi_{\mathrm{deloc}}^{(0)}\,,
\end{equation}
with $\Psi_{\mathrm{loc},\mathrm{deloc}}^{(0)}$ linear combinations of
localized or delocalized unperturbed modes only, with norm of order
1. (If the contribution of delocalized unperturbed modes comes with a
higher power of $g$, the bound is ineffective.)
This seems unlikely: in fact, matrix elements $H_{n'n}$ are of order
$O(g)\cdot O(1)$ only for pairs of localized modes with close or
overlapping spatial support, which on the other hand are typically
well separated in the spectrum. One then expects their mixing to be
negligible when computing the exact eigenmodes, and so one expects
to find localized exact modes only, if only localized unperturbed
modes are involved.

To see that this is the case, we impose that $\psi_n$ is an exact
eigenvector with eigenvalue $E$, and
projecting on the localized unperturbed mode $\psi_{n}^{(0)}$, we
obtain the equation
\begin{equation}
  \label{eq:pert_equ0}
\begin{aligned}
E c_n &= 
E_{n}^{(0)} c_n+
g\left(\sum_{n'\in\mathrm{loc}}\Delta_{nn'} c_{n'}\right) + g^{1+\epsilon}r_n
\end{aligned}
\end{equation}
where $c_n= (\psi_{n}^{(0)},\Psi_{\mathrm{loc}}^{(0)})$ and
$r_n=(\psi_{n}^{(0)},\Delta\Psi_{\mathrm{deloc}}^{(0)})$. Since $g$ is
small, we can attempt to solve it perturbatively, checking afterward
that the resulting mixing coefficients are small.
Setting $E=E_0 + t E_1 + \ldots$ and $c_n= c_n^{(0)} +
tc_n^{(1)}+\ldots$, and replacing $g\Delta\to t g\Delta$ and $r_n\to t
r_n$,  with $t$ a formal expansion parameter that will
be eventually set to 1,  one finds
\begin{equation}
  \label{eq:pert_equ1}
  \begin{aligned}
    &[E_0 - E_n^{(0)} + t(E_1- g\Delta_{nn})+O(t^2)](c_n^{(0)} + t c_n^{(1)}+O(t^2))
    \\ & = tg\sum_{\substack{n'\in\mathrm{loc}\\n'\neq n}} \Delta_{nn'} 
(    c_{n'}^{(0)} + t    c_{n'}^{(1)}+O(t^2)) + t^2g^{1+\epsilon} r_n
\end{aligned}
\end{equation}
yielding to lowest orders in $t$ 
the equations
\begin{equation}
  \label{eq:pert_equ2}
  \begin{aligned}
  (E - E_n^{(0)})c_n^{(0)}  &= 0\,,\\
  (E_0 - E_n^{(0)})c_n^{(1)} +  (E_1- g\Delta_{nn})c_n^{(0)}  &=
  \sum_{\substack{n'\in\mathrm{loc}\\n'\neq n}} g\Delta_{nn'} c_{n'}^{(0)}\,.
\end{aligned}
\end{equation}
Notice that the contribution of delocalized unperturbed modes plays no
role here.  The first equation tells us that to leading order,
$E=E_{\tilde{n}}^{(0)}$ for some $\tilde{n}$, with nonzero $c_n^{(0)}$
only in the corresponding degenerate unperturbed subspace. 
We then choose one of the four unperturbed basis vectors and set
$c_{\tilde{n}}^{(0)}=0$ and $c_n^{(0)}=0$ if $n\neq\tilde{n}$; the
other three choices can be treated analogously. The second equation then tells us that
$E_1=g\Delta_{\tilde{n}\tilde{n}}$, and for $n$
such that $E_n^{(0)}\neq E_{\tilde{n}}^{(0)}$, it 
tells us that
\begin{equation}
  \label{eq:pert_equ3}
c_n^{(1)} = \f{g\Delta_{n\tilde{n}}}{E_{\tilde{n}}^{(0)}-E_n^{(0)}}\,.
\end{equation}
These coefficients are generally 
\begin{equation}
  \label{eq:pt9}
c_n^{(1)} 
  =
  g\f{O(e^{-d_{n'n}/\xi_{n'n}})}{O(1)}\,,
\end{equation}
if the unperturbed eigenvalues are not close, and only for order
$O(1)$ modes (i.e., those with small $d_{n'n}/\xi_{n'n}$) one finds a
numerator of order $O(1)$. If the unperturbed eigenvalues are nearby
in the spectrum, i.e., they differ by $O(1/V_\Lambda)$, then there is
only a $O(1/V_\Lambda)$ chance that also the corresponding
eigenvectors are localized near each other in space, and so for the
typical coefficient one has the estimate\footnote{For a more
    accurate estimate when $g>1/V_\Lambda$, spatially close modes with
    exactly or nearly degenerate (i.e.,
    $E_n^{(0)}-E_{n'}^{(0)}=O(1/V_\Lambda)$) eigenvalues should be
    treated together as degenerate modes, putting any small difference
    into the perturbation.
    In this case the magnitude of the perturbation (after including in
    it the small eigenvalue differences) remains of order $O(g)$.
    The perturbation should then be diagonalized exactly in the
    corresponding subspace, which has dimension $O(1)$, since the
    probability of finding $k$ modes nearby both in space and in the
    spectrum is suppressed as $1/V_\Lambda^k$.
    The mixing of nearby localized modes leads, of course, to
    similarly localized modes. For modes in different subspaces, which
    are now spectrally well-separated, the perturbative analysis
    discussed above works without any issue. Further mixing of modes
    within a degenerate subspace cannot change their localization properties,
    see also discussion below.}
\begin{equation}
  \label{eq:pt10}
  c_n^{(1)}   =
  \f{O(g/V_\Lambda)}{O(1/V_\Lambda)} = O(g)\,,
\end{equation}
independently of the volume, which is small, as it should be for
perturbation theory to be valid, and negligible compared to
$c_{\tilde{n}}\simeq c_{\tilde{n}}^{(0)}=1$. This surely does not lead
to delocalization, and changes only marginally the size of the modes.
While $\Re c_{\tilde{n}}^{(1)}=0$ due to the normalization condition,
and $\Im c_{\tilde{n}}^{(1)}=0$ with a suitable choice of phases,
$c_{n}^{(1)}$ for $E_n^{(0)}= E_{\tilde{n}}^{(0)}$ but
$n\neq\tilde{n}$ are not determined at this stage. To do that we write
Eq.~\eqref{eq:pert_equ1}, which in this case 
starts from $O(t^2)$, and reads
\begin{equation}
  \label{eq:pert_equ1_deg}
  \begin{aligned}
    & t^2g(\Delta_{\tilde{n}\tilde{n}}- \Delta_{nn}) c_n^{(1)} + O(t^3)
    \\ &
     = t^2\left(\sum_{\substack{n'\in\mathrm{loc}\\n'\neq n,\tilde{n}}}
       g\Delta_{nn'}     c_{n'}^{(1)}\right) + t^2 g^{1+\epsilon} r_n
     + O(t^3)\,,
 \end{aligned}
\end{equation}
yielding (for $E_n^{(0)}= E_{\tilde{n}}^{(0)}$ but
$n\neq\tilde{n}$)
\begin{equation}
  \label{eq:pert_equ1_deg2}
  \begin{aligned}
  c_n^{(1)}
&  = \f{\sum_{\substack{n'\in\mathrm{loc}\\n'\neq n,\tilde{n}}}
  \Delta_{nn'}     c_{n'}^{(1)}}{\Delta_{\tilde{n}\tilde{n}}-
  \Delta_{nn} } + g^\epsilon \f{r_n}{\Delta_{\tilde{n}\tilde{n}}-
  \Delta_{nn} }    \\
&  = g\sum_{\substack{n'\in\mathrm{loc}\\n'\neq n,\tilde{n}}}\f{
  \Delta_{nn'}\Delta_{n'\tilde{n}} }{(\Delta_{\tilde{n}\tilde{n}}-
  \Delta_{nn})(E_{\tilde{n}}^{(0)}-E_{n'}^{(0)}) } \\ 
  &\phantom{=}+ g^\epsilon \f{r_n}{\Delta_{\tilde{n}\tilde{n}}- \Delta_{nn} }    \,,
  \end{aligned}
\end{equation}
having assumed full lifting of the degeneracy (i.e.,
$\Delta_{nn}\neq \Delta_{\tilde{n}\tilde{n}}$ if
$E_n^{(0)}=E_{\tilde{n}}^{(0)}$ and $n\neq \tilde{n}$). 
For $\epsilon<1$, the second term dominates, 
but it still gives a small contribution when $g\ll 1$, so perturbation
theory can be trusted.\footnote{For the leading coefficient
    $c_{\tilde{n}}$, the normalization condition
    $\Vert\psi_n\Vert^2=1$, together with
    Eqs.~\eqref{eq:pert_delovec}, \eqref{eq:pert_equ3},
    \eqref{eq:pert_equ1_deg2}, and the fact that only $O(1)$ of the
    $c_n$ with $n\neq \tilde{n}$ are non-negligible, implies
    $c_{\tilde{n}}= 1 + O(g^{2\epsilon})$.} (The fact that the
solution is not analytic in $g$ is irrelevant.) Furthermore, mixing
between degenerate modes, no matter how large, cannot lead to
delocalization, since these modes have the same localization
properties and similar spatial support.
To fully determine the eigenvector $\psi_n$ one would
need to determine $\Psi^{(0)}_{\mathrm{deloc}}$ self-consistently, but
this is not relevant for our goal: for us, it is enough having shown
that a delocalized mode of the required type cannot be found, no
matter what $\Psi^{(0)}_{\mathrm{deloc}}$ is. 

While one generally expects perturbation theory to break down in a
large volume, in this case it does not.  The reason is that the number
of localized modes with non-negligible spatial overlap remains always
of order $O(1)$, no matter how large the volume is; and that even
though the level spacing becomes of order $O(1/V_\Lambda)$,
neighboring modes in the spectrum have only a chance of order
$O(1/V_\Lambda)$ to be also close in space (and so to have
non-negligible overlap). The two factors essentially cancel out, and
mixing still involves $O(1)$ modes with a factor of
order $g$, which is small in our setup. In other words, mixing of an
unperturbed localized mode with different unperturbed localized modes
will make up a tiny fraction of the perturbed mode's weight, and so
cannot change its localized nature.

In conclusion, no delocalized mode as in 
Eq.~\eqref{eq:pert_delovec} can be built out of localized modes alone,
showing that if the exact mode $\psi_n$ is delocalized, then
$E_n> E_c^{(0)} - \delta_-$, with $\delta_-$ a positive number of
order $O(g^\epsilon)$, for any $0<\epsilon\le 1$. (Notice that the
case $\epsilon=0$ is excluded, so that delocalized exact modes
including an $O(1)$ contribution from delocalized unperturbed modes
can exist. These, however, cannot be found farther away than $O(g)$
from the unperturbed mobility edge.)  Together with the upper bound
found above, this leads us to conclude that for the exact mobility
edge $E_c$ one has $|E_c - E_c^{(0)}|=O(g)$.

All in all, the possible effect of taste-breaking interactions on the
mobility edge (in a finite volume) is to shift its unperturbed value
by at most $O(g)$.  As these are short-ranged, UV effects, there are
no further volume dependencies expected. 
Besides taste-breaking effects, other finite-spacing
effects amount only to renormalizing the whole spectrum like a quark
mass~\cite{DelDebbio:2005qa,Giusti:2008vb,Bonanno:2019xhg,
  Bonanno:2023ypf,Bonanno:2023xkg}.

The argument above shows that when $g<1/V_\Lambda$, changes in the
eigenvalues of would-be degenerate localized modes induced by
introducing a localized fluctuation in the gauge field will be
strongly correlated, as they are mainly driven by the $O(1/V_\Lambda)$
change in the corresponding unperturbed eigenvalue.  This gives a
microscopic explanation for the change in local spectral statistics in
the presence of an approximate taste symmetry, discussed in the
previous subsection. On the other hand, while the approximate taste
symmetry strongly affects the determination of the mobility edge from
spectral statistics, it does not affect much the localization
properties of modes and the true position of the mobility edge, even
in this regime.

As already mentioned above, on top of finite-spacing effects there are
finite-size effects that can shift the finite-volume estimate of the
mobility edge by at most $O(1/L_\Lambda^{1/\nu})$. This can occur
because of mixing of localized and delocalized modes around the
mobility edge, where the size of localized modes is of the order of
the volume of the system.  These are IR effects controlled by the
ratio of the localization/correlation length of eigenmodes and the
linear size of the system, which is a ratio of physical scales and, as
such, a renormalization-group-invariant quantity, that will become
independent of the spacing as this goes to zero. That the correlation
length of eigenvectors is physical follows from the fact that
correlators such as
$\la \sum_n \delta(\lambda -\lambda_n)\Vert\psi_n(x)\Vert^2
\Vert\psi_n(0)\Vert^2\ra $ renormalize multiplicatively, up to
renormalizing also the spectrum~\cite{Giordano:2022ghy}.

In conclusion, returning to physical units and renormalizing the
spectrum, for the exact and unperturbed mobility edges in
quark mass units,
$\bar{\lambda}_{c} = \lambda_c/m_s = \Lambda E_c/m_s$ and
$\bar{\lambda}_{c}^{(0)} = \lambda_c^{(0)}/m_s = \Lambda
  E_c^{(0)}/m_s$, one has
\begin{equation}
  \label{eq:pt13}
  \begin{aligned}
    \bar{\lambda}_{c}&= 
    \bar{\lambda}_{c}^{(0)}+ C_{\text{finite-size}}(\Lambda
    L)^{-\f{1}{\nu}} \\ &\phantom{=} +
    C_{\text{taste-breaking}}(a\Lambda)^2 + \ldots\,,
  \end{aligned}
\end{equation}
where for the behavior of finite-size corrections we conservatively
used its expected upper bound. Notice that taking the thermodynamic limit
first one finds
$\lim_{L\to \infty} \bar{\lambda}_{c}-\bar{\lambda}_{c}^{(0)}
= O\left((a\Lambda)^2\right)$.
Since $\bar{\lambda}_{c}^{(0)}$ is by construction not affected by taste-breaking
effects, one expects that continuum and thermodynamic limits commute for it, i.e.,
\begin{equation}
  \label{eq:pt15}
  \bar{\lambda}_{c}^{(0)}(a,L) =
  \bar{\lambda}_{c}^{(0)}(0,\infty) + \ldots\,,
\end{equation}
where corrections, vanishing in the continuum and thermodynamic limit
taken in any order, are expected to be at most $O(1/L^{1/\nu})$ in the
system size, and likely smaller, and $O\left((a\Lambda)^2\right)$ in
the lattice spacing thanks to the lattice chiral symmetry of staggered
fermions.  Then Eq.~\eqref{eq:pt13} shows that the two limits commute
also for $\lambda_c/m_s$.

\section{Numerical results}
\label{sec:numres}

In this section, after briefly summarizing our lattice setup, we
present our results for the mobility edge, discussing in detail the
effects of taste degeneracy on the spectral statistics.

\subsection{Lattice setup}
\label{sec:numressetup}

In this work we considered 5 ensembles of gauge configurations of
$N_f=2+1$ QCD at the physical point, originally generated for the
investigation of the QCD topological susceptibility at finite
temperature reported in Ref.~\cite{Athenodorou:2022aay}, obtained on
$N_s^3 \times N_t$ lattices with lattice spacings $a$ ranging from
$\sim 0.107\,\mathrm{fm}$ down to $\sim 0.054\,\mathrm{fm}$. In all
cases, the temperature was fixed to
$T=1/(aN_t) \simeq 230\,\mathrm{MeV}$, and the lattice size to
$L = a N_s \simeq 3.4\,\mathrm{fm}$, corresponding to an aspect ratio
$N_s/N_t=LT=4$. For the next-to-finest lattice spacing, corresponding
to $N_t=14$, we also considered 2 additional ensembles with different
lattice volumes, with aspect ratios $LT\simeq 3.4$ and $\simeq 4.6$,
corresponding respectively to $L \simeq 2.9\,\mathrm{fm}$ and
$\simeq 3.9\,\mathrm{fm}$. In order to keep our paper self-contained,
we briefly summarize below the lattice setup employed in
Ref.~\cite{Athenodorou:2022aay} to generate configurations, and we
refer the reader to the original reference for further technical
details.

Gauge configurations were generated adopting the tree-level Symanzik
improved Wilson gauge action~\cite{Weisz:1982zw,Curci:1983an,
  Symanzik:1983dc,Luscher:1984xn} to discretize the pure-Yang--Mills
term, and rooted stout-smeared staggered fermions to discretize the
Dirac term. Expectation values are then defined schematically as
\begin{equation}
  \label{eq:expval_def}
  \la \Oc\ra = \f{\int [dU]\,
    e^{-S_{\mathrm{g}}[U;\beta]-S_{\mathrm{f,eff}}[U;m_l,m_s]}
    \Oc[U]}{\int
    [dU]\,    e^{-S_{\mathrm{g}}[U;\beta]-S_{\mathrm{f,eff}}[U;m_l,m_s]}}\,.
\end{equation}
Here $S_g$ denotes the discretized pure-Yang--Mills term, whose exact
form is not relevant; $U_\mu(x)\in \SU(3)$ is the gauge link variable
attached to the link connecting the sites $x$ and $x+a\hat{\mu}$, and
$[dU]=\prod_{x,\mu}dU_\mu(x)$ is the product of the corresponding Haar
integration measures. Moreover,
\begin{equation}
  \label{eq:expval_def2}
  \begin{aligned}
    e^{-S_{\mathrm{f,eff}}[U;m_l,m_s]}&=M[U^{(2)};m_l]^{\f{1}{2}}M[U^{(2)};m_s]^{\f{1}{4}}\,,\\
    M[U^{(2)};m] &= \det \left(D^{\mathrm{stag}}[U^{(2)}] + m\right)
    \\
    &= \prod_n \left(i\lambda_n[U^{(2)}] + m\right)\,,
  \end{aligned}
\end{equation}
which is real positive thanks to the symmetry of the spectrum [see
after Eq.~\eqref{eq:stag_op2}].  Here $U_\mu^{(2)}(x)\in \SU(3)$ is
obtained from $U_\mu(x)$ after 2 steps of isotropic stout
smearing~\cite{Morningstar:2003gk} with smearing parameter
$\rho = 0.15$.  The bare (inverse) coupling $\beta$, the bare light
and strange quark masses $m_l$ and $m_s$, and the lattice spacing $a$
were fixed in Ref.~\cite{Athenodorou:2022aay} according to the results
of Refs.~\cite{Aoki:2009sc, Borsanyi:2010cj, Borsanyi:2013bia} so as
to stay at the physical point, defined by a Line of Constant Physics
with the pion mass $m_\pi = 135\,\mathrm{MeV}$ and the
strange-to-light quark mass ratio $R=m_s/m_l=28.15$ fixed to their
physical values. All simulation parameters are summarized in
Tab.~\ref{tab:simulation_params}.

The Monte Carlo simulations of Ref.~\cite{Athenodorou:2022aay}
employed the Rational Hybrid Monte Carlo (RHMC) updating
algorithm~\cite{Clark:2004cp,Clark:2006fx,Clark:2006wp}, used, for
some simulation points, in combination with the \textit{multicanonical
  algorithm}~\cite{Berg:1992qua,Bonati:2017woi,Jahn:2018dke,
  Bonati:2018blm,Athenodorou:2022aay,Bonanno:2022dru}. As a matter of
fact, it is well known that, above the QCD chiral crossover
$T_c\simeq 155\,\mathrm{MeV}$, the topological susceptibility
$\chi=\la Q^2\ra/V_4$, where $V_4=L^3/T$, is rapidly suppressed as a
function of the temperature~\cite{RevModPhys.53.43,Borsanyi:2015cka,
Petreczky:2016vrs,Borsanyi:2016ksw,Lombardo:2020bvn,Borsanyi:2021gqg,
Athenodorou:2022aay}, leading to $\la Q^2\ra = V_4 \chi \ll 1$ for the 
lattices employed in typical simulations. Multicanonical simulations are
employed to enhance the probability of visiting nontrivial topological
sectors when this probability is suppressed, without spoiling
importance sampling. For what concerns the present investigation, the
multicanonical algorithm allows one to avoid possible systematic
effects in the low-lying staggered spectrum, which is connected to
topological excitations. Concerning the technical details of
multicanonical runs, we refer the reader to the original
work~\cite{Athenodorou:2022aay}.

\begin{table}[!t]
\begin{center}
\begin{tabular}{|c|c|c|c|c|c|}
\hline
\multicolumn{6}{|c|}{}\\[-1em]
\multicolumn{6}{|c|}{$T=230$ MeV, $\quad T/T_c=1.48$}\\
\hline
&&&&&\\[-1em]
$\beta$ & $a$~[fm] & $am_s \cdot 10^{-2}$ & $N_s^3 \times N_t$ & $L$~[fm] & Statistics\\
\hline
&&&&&\\[-1em]
3.814 & 0.1073 & 4.27 & $32^3 \times 8$* & 3.43 & 1025 \\
\hline
&&&&&\\[-1em]
3.918 & 0.0857 & 3.43 & $40^3 \times 10$* & 3.43 & 595 \\
\hline
&&&&&\\[-1em]
4.014  & 0.0715 & 2.83 & $48^3 \times 12$ & 3.43 & 624 \\
\hline
&&&&&\\[-1em]
\multirow{3}{*}{4.100}  & \multirow{3}{*}{0.0613} & \multirow{3}{*}{2.40} & $48^3 \times 14$ & 2.94 & 224 \\
&&& $56^3 \times 14$  & 3.43 & 780 \\
&&& $64^3 \times 14$* & 3.92 & 228 \\
\hline
&&&&&\\[-1em]
4.181  & 0.0536 & 2.10 & $64^3 \times 16$ & 3.43 & 320 \\
\hline
\end{tabular}
\end{center}
\caption{Simulation parameters used in
  Ref.~\cite{Athenodorou:2022aay} to generate the
  physical-point gauge ensembles considered in this work. The bare
  parameters $\beta$, $am_s$, $am_l = am_s / 28.15$ and the lattice
  spacing $a$ have been fixed according to the results of
  Refs.~\cite{Aoki:2009sc, Borsanyi:2010cj, Borsanyi:2013bia}, and
  correspond to physical pion mass and physical strange-to-light quark
  mass ratio. Simulations marked with * have been performed without
  the multicanonical algorithm.}
\label{tab:simulation_params}
\end{table}

\begin{figure}[!t]
\includegraphics[scale=0.38]{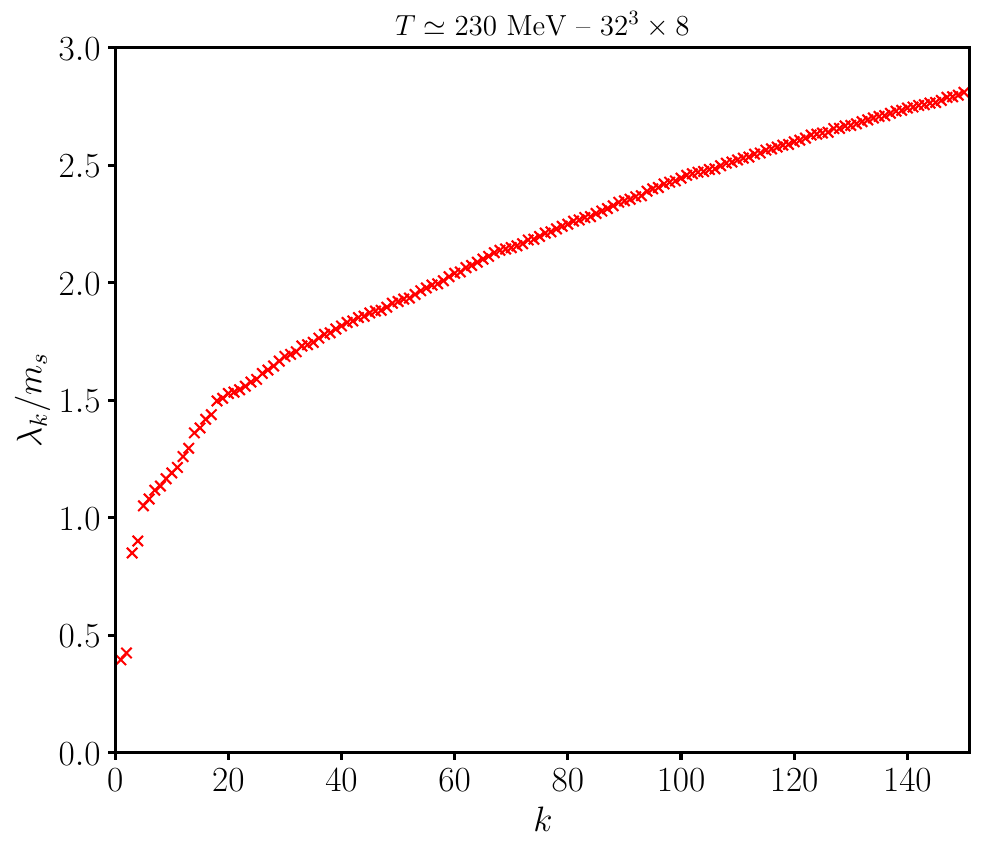}
\caption{Eigenvalues $\lambda_k/m_s$ of the staggered operator in the
  background of a typical gauge configuration on a $32^3 \times 8$
  lattice at $T=230\,\mathrm{MeV}$.}
\label{fig:spectrum_ex}
\end{figure}

For each gauge configuration we then analyzed the lowest 150 positive
eigenvalues of $D^{\mathrm{stag}}[U^{(2)}]$ [i.e., the same operator
appearing in the fermionic determinant, Eq.~\eqref{eq:expval_def2}],
obtained in Ref.~\cite{Athenodorou:2022aay} using the PARPACK
library~\cite{PARPACK}, from which we computed the corresponding
unfolded eigenvalues. We complemented these data by performing a new
calculation to obtain the IPRs of the Dirac eigenvectors for the gauge
ensembles with $N_t=14$. We calculated error bars from a standard binned 
jack-knife performed over the whole analysis, 
including the unfolding procedure (when studying spectral statistics), 
to take into account possible correlations introduced by the latter.
An example of a staggered Dirac spectrum on a
single, typical configuration is shown in Fig.~\ref{fig:spectrum_ex}.
We performed unfolding by ordering all the available eigenvalues
$\lambda_n$ of all the configurations of a given lattice ensemble, and
replacing them with their rank divided by the number of
configurations~\cite{Kovacs:2010wx}. This automatically yields unit
spectral density, while the deviation of $\la s\ra$ from 1 can be used
to test the accuracy of the procedure.

To compute $I_{s_0}$ and $\Var(s)$ locally in the spectrum we
approximated Eq.~\eqref{eq:unfold2} by binning the spectrum in bins of
size $0.178 m_s$ in physical units, averaging inside each bin, and
assigning the result to the center of the bin.  Since, loosely
speaking, the Dirac spectrum renormalizes like the quark
mass~\cite{DelDebbio:2005qa,Giusti:2008vb,Bonanno:2019xhg,
  Bonanno:2023ypf,Bonanno:2023xkg}, keeping the bin size fixed in
units of $m_s$ is the appropriate choice.  An unfolded spacing was
included in a bin average if the corresponding lowest (not unfolded)
eigenvalue fell in the bin.

\begin{figure}[!t]
\centering
\includegraphics[scale=0.4]{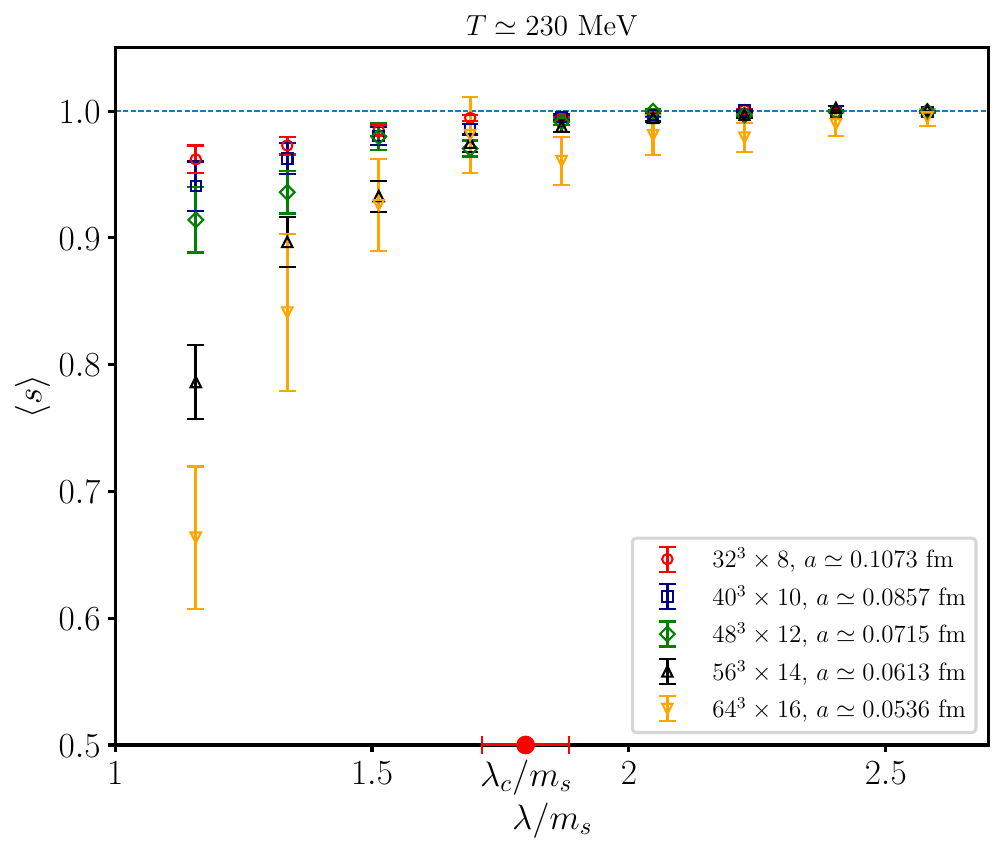}
\includegraphics[scale=0.405]{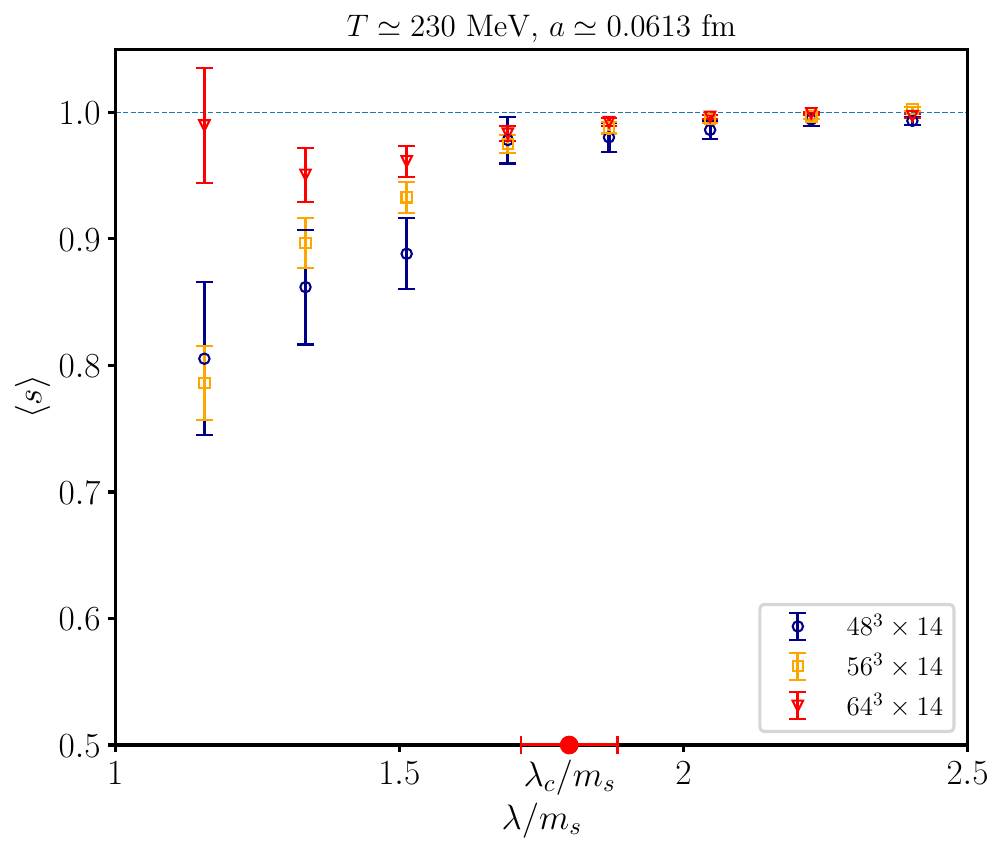}
\caption{Top panel: comparison of $\la s \ra$ as a function of
  $\lambda/m_s$ for all the explored lattice spacings at fixed value
  of the aspect ratio $LT=4$. Bottom panel: comparison of $\la s \ra$
  as a function of $\lambda/m_s$ for all the explored aspect ratios,
  corresponding to different values of $N_s$, at fixed value of the
  lattice spacing, corresponding to a temporal extent $N_t=14$. Our
  continuum estimate for the mobility edge [see  
  Eq.~\eqref{eq:cont_lim_ave_comb} in Sec.~\ref{sec:lc_cont_final}] 
  is marked in both panels by a red circle on the horizontal axis.}
\label{fig:s_mean_comp}
\end{figure}

We similarly computed $\la s\ra$, and found it compatible with 1
within errors in the relevant spectral region $\lambda/m_s\ge 1.5$,
see Fig.~\ref{fig:s_mean_comp}. This reassures us on the validity of
our unfolding procedure.  Between 1 and 1.5 there is a small but
significant deviation from 1, systematically increasing as one goes
down in $\lambda$. This is well understood~\cite{Giordano:2019pvc},
and it is due simply to the low but rapidly changing spectral density,
that requires the use of large bins to have a decent signal, at the
price of having a non-constant density inside the bin. This leads to
the smaller eigenvalue spacings corresponding to modes at the higher,
and denser, end of the bin, lowering the numerical estimate
$\la\Delta\lambda\ra_{\mathrm{bin}}$ of the local average spacing
between neighboring eigenvalues in a spectral bin below the expected
value $1/\rho_{0\,\mathrm{bin}}$, leading to
$\la s\ra_{\mathrm{bin}}\simeq \la\Delta\lambda\ra_{\mathrm{bin}}\,
\rho_{0\,\mathrm{bin}} < 1$ for the numerical estimate of $\la
s\ra$. Between 1 and 1.5 one can also see a systematically increasing
deviation from 1 as $N_t$ increases (see Fig.~\ref{fig:s_mean_comp},
top panel). This is due to the formation of distinct taste-degenerate
multiplets in this spectral region which, being most likely found at
the higher end of a spectral bin, are also more likely to spread
across neighboring bins. This leads to lowering
$\la\Delta\lambda\ra_{\mathrm{bin}}$ further, as spacings within a
taste multiplet are more likely to contribute than spacings between
multiplets.  This explanation is supported by the fact that increasing
$N_s$ at fixed $N_t$, so bringing the multiplets closer, reduces the
distance of $\la s\ra_{\mathrm{bin}}$ from 1, see
Fig.~\ref{fig:s_mean_comp}, bottom panel. Finally, for
$\lambda/m_s<1$ the spectrum is very sparse, and it is hard to make
any reliable statement.

\begin{figure}[!t]
\includegraphics[scale=0.39]{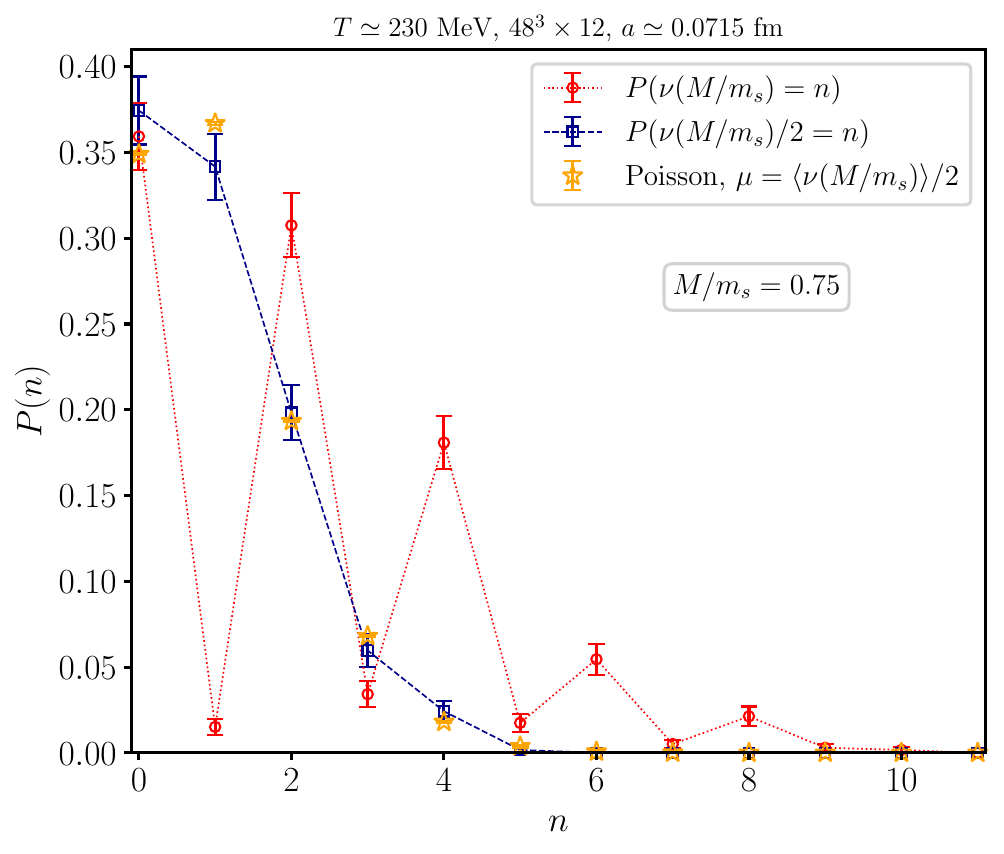}
\caption{Mode counting in the localized regime of the spectrum.  Red
  circles show the probability $ P(\nu(M/m_s)=n)$ of finding $n$ modes
  below the cutoff $M=0.75 m_s$, deep in the localized regime. Blue
  squares show the probability $ P(\nu(M/m_s)/2=n)$ of finding $n$
  mode doublets below the cutoff $M$. Yellow stars show the
  corresponding Poisson distribution with parameter
  $\mu = \la\nu(M/m_s)/2\ra$.}
\label{fig:mode_count_distr}
\end{figure}

\subsection{Effects of taste degeneracy on mode countings}

As a preliminary piece of evidence of the strong effects of taste
degeneracy on the statistical properties of the spectrum, in
Fig.~\ref{fig:mode_count_distr} we show with red circles the
probability distribution of the number $\nu(M/m_s)$ of eigenmodes
found below a fixed cutoff, i.e., $\lambda_n\le M$, with $M$ chosen
safely below the mobility edge (see below Sec.~\ref{sec:lc_cont_final}).
Notice that $\nu$ is a renormalized quantity if the renormalized
physical value of $M$ is kept fixed~\cite{DelDebbio:2005qa,
  Giusti:2008vb,Bonanno:2019xhg,Bonanno:2023ypf,Bonanno:2023xkg},
e.g., if $M/m_s$ is kept fixed.  While the expectation is that this
counting follows a Poisson distribution with parameter equal to the
average number of modes below the cutoff, $\la\nu(M/m_s)\ra$, the data
show otherwise.

Quite striking at first sight is the oscillating behavior that leaves
the odd bins almost empty. This, however, is easily understood if we
take notice that the low-lying spectrum shows the formation of
eigenvalue doublets (see Fig.~\ref{fig:spectrum_ex}), the first step
in the formation of the quartets expected in the continuum limit (see,
e.g., Ref.~\cite{Fodor:2009wk}). If we correct for this by counting
the number of doublets, $\nu/2$, instead of the number of modes, and
compare with the Poisson distribution with parameter equal to the
average number of doublets (i.e., half the average number of modes)
below the cutoff, then we find excellent agreement between the two
curves, shown with blue squares and yellow triangles in
Fig.~\ref{fig:mode_count_distr}.

This is a simple but convincing demonstration that spectral statistics
are heavily distorted by taste-degeneracy effects on fine lattices. It
also shows that the taste doublets of eigenmodes fluctuate
independently, as expected for localized modes in the absence of
near-degeneracy. We now proceed with a quantitative assessment of
these effects on the unfolded level spacing distribution.

\subsection{Determination of the mobility edge from spectral
  statistics}

We estimated the mobility edge using spectral statistics as the point
where these match their critical behavior using two features of the
unfolded level spacing distribution, namely $I_{s_0}$ and
$\Var(s)$. For $I_{s_0}$, the critical value was obtained in
Ref.~\cite{Giordano:2013taa} by means of a finite-size-scaling
analysis, and reads
\begin{equation}
  \label{eq:is0crit}
I_{s_0,c}=0.1966(25)\,.  
\end{equation}
Here we determined the critical value of $\Var(s)$ by means of a
similar analysis on the same lattice data ($2+1$ QCD at $T\simeq 2.6T_c$,
$a\simeq 0.125\,\mathrm{fm}$, physical $m_{l,s}$):\footnote{We thank
  T.G.~Kov\'acs and F.~Pittler for allowing us to use the data.}
technical details about the procedure can be found in
Ref.~\cite{Giordano:2013taa}. Our estimate is
\begin{equation}
  \label{eq:varcrit}
\Var(s)_c= 0.3702(98)\,.  
\end{equation}
The central value is obtained using only lattices of spatial size
$N_s\ge 40$ and data points in a range of width
$a\Delta \lambda = 0.026$ around the mobility edge. The error includes
the contribution of the statistical error from the fit, performed with
the MINUIT routine~\cite{James:1975dr}; the systematic finite-volume
effect estimated as the change of the fit result due to including also
$N_s=36$ data; and the systematic of the fitting range estimated as
the change of the fit result due to shifting the fitting range down in
the spectrum by $10\%$ of its width.\footnote{We also find
  $a\lambda_c^{(T=2.6T_c)} = 0.33602(63)$ for the mobility edge and
  $\nu = 1.406(98)$ for the localization-length critical exponent, in
  agreement with those found in Ref.~\cite{Giordano:2013taa}.}

\begin{figure}[!t]
\centering
\includegraphics[scale=0.41]{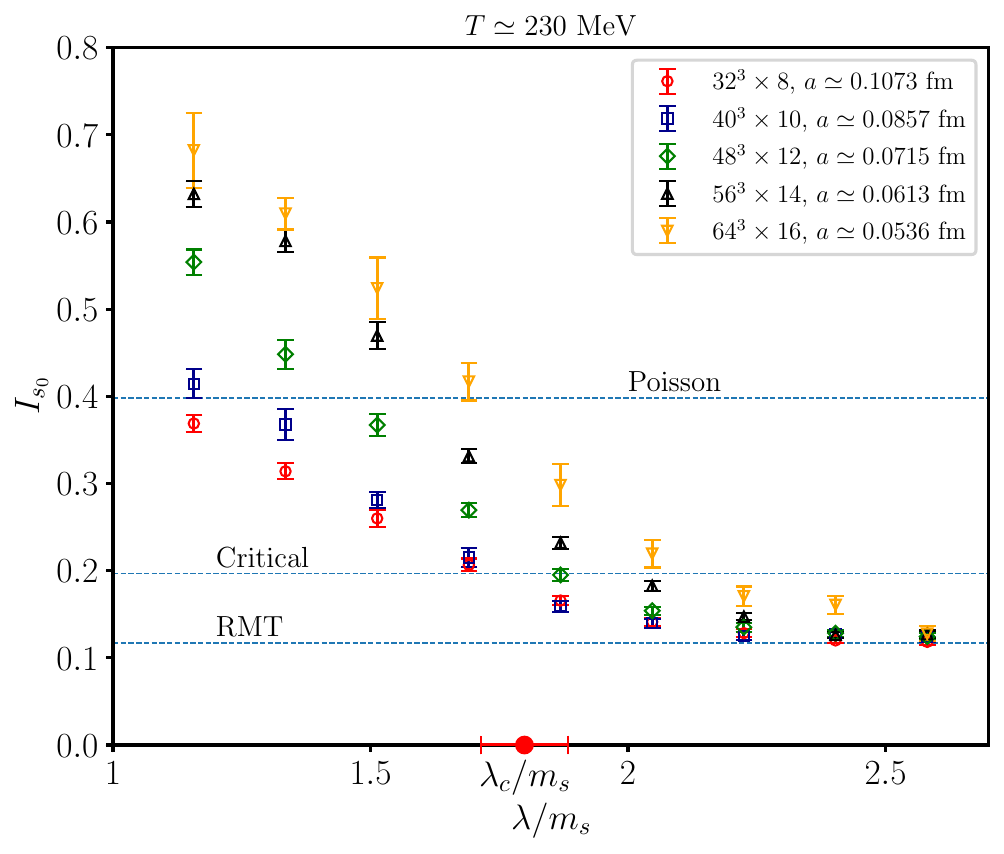}
\includegraphics[scale=0.41]{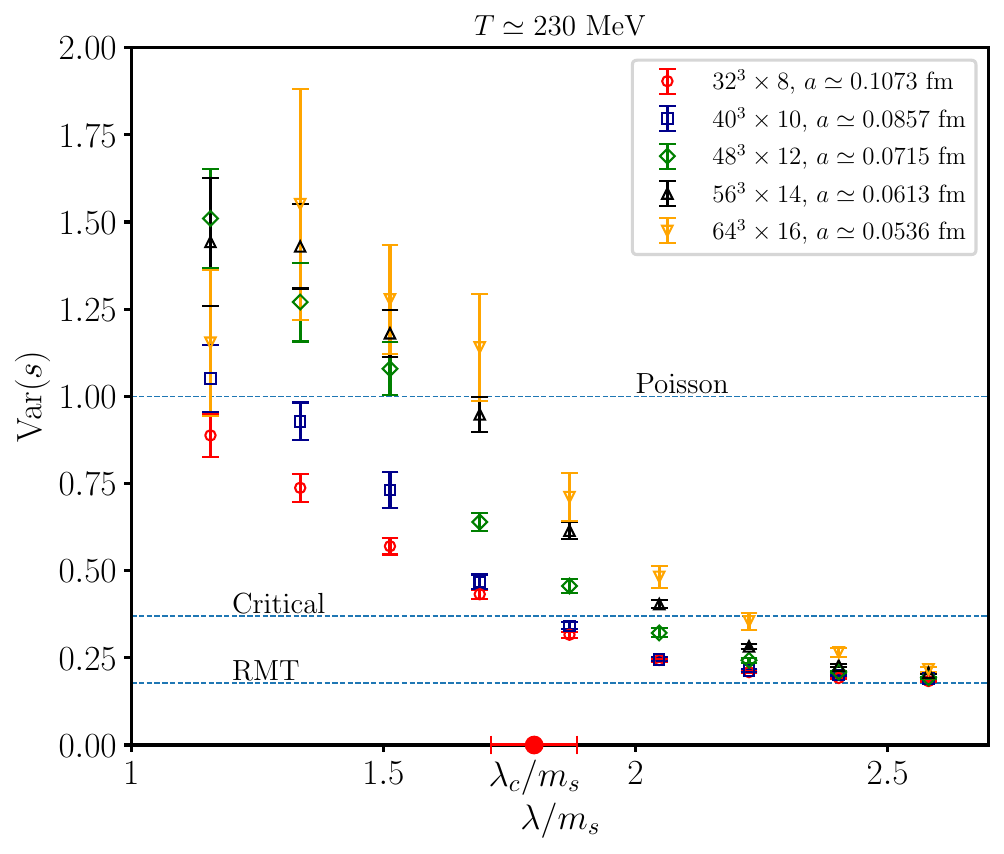}
\caption{Comparison of $I_{s_0}$ (top panel) and $\Var(s)$ (bottom
  panel), computed locally in the spectrum on lattices at fixed
  temperature $T=230\,\mathrm{MeV}$, for different values of the
  lattice spacing at fixed spatial lattice size $L=3.4\,\mathrm{fm}$
  in physical units (i.e., fixed $LT=4$). Our continuum estimate for
  the mobility edge 
  [see Eq.~\eqref{eq:cont_lim_ave_comb} in 
  Sec.~\ref{sec:lc_cont_final}] is marked by a
  red circle on the horizontal axis.}
\label{fig:mob_edge_COMP}
\end{figure}

\begin{figure}[!t]
  \centering
  \includegraphics[scale=0.41]{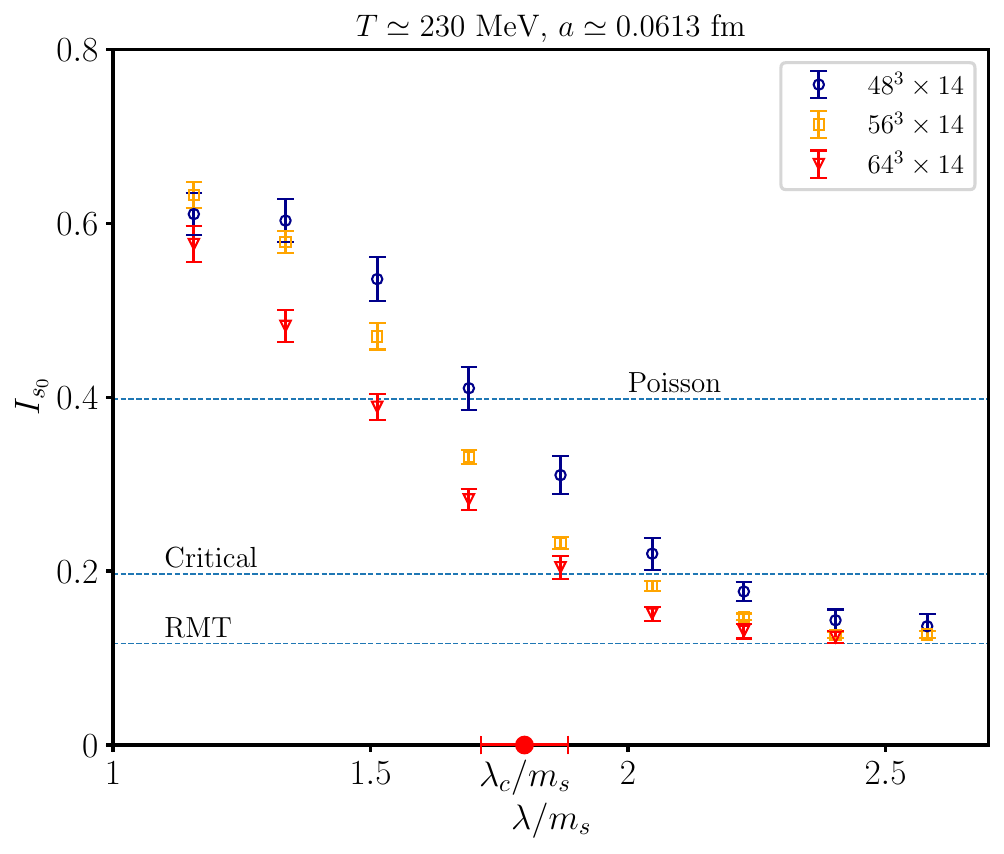}
  \includegraphics[scale=0.41]{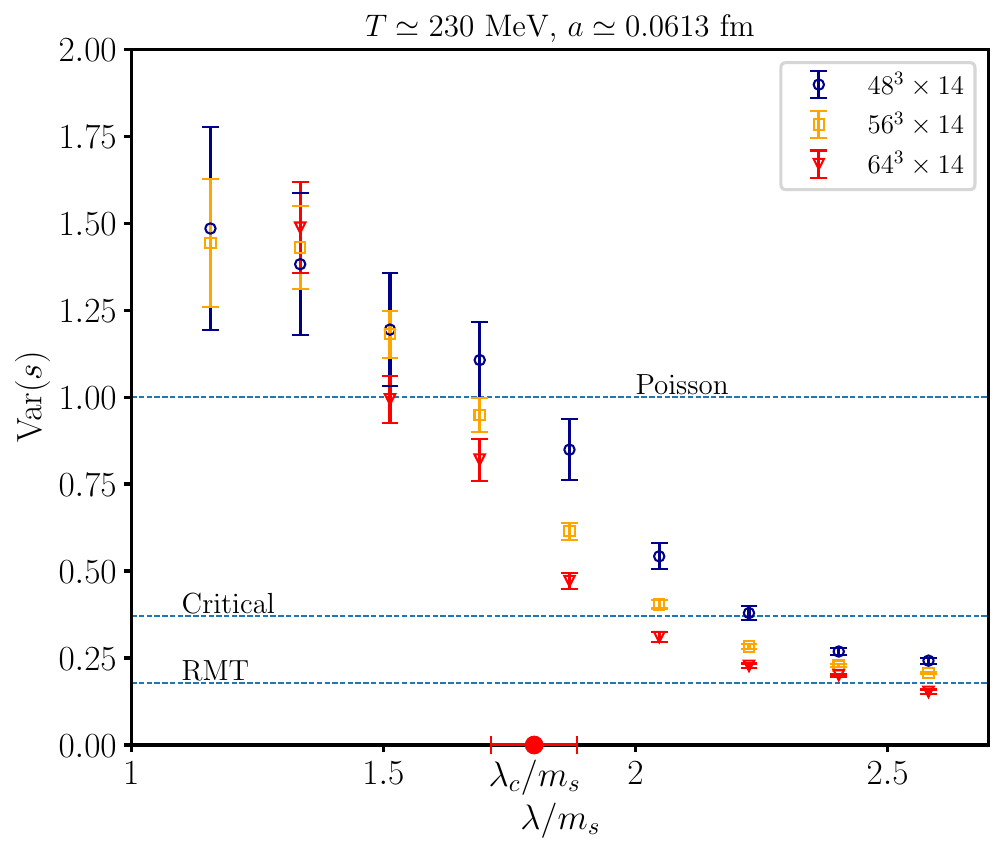}
  \caption{Comparison of $I_{s_0}$ (top panel) and $\Var(s)$ (bottom
    panel), computed locally in the spectrum on lattices at fixed
    temperature $T=230\,\mathrm{MeV}$, for different spatial volumes
    at fixed lattice spacing, $a=0.0615\,\mathrm{fm}$ (i.e.,
    $N_t=14$). Our continuum estimate for the mobility edge 
    [see Eq.~\eqref{eq:cont_lim_ave_comb} in Sec.~\ref{sec:lc_cont_final}]
    is marked by a red circle on the
    horizontal axis.}
\label{fig:mob_edge_COMP_Ns}
\end{figure}

\begin{figure}[!t]
  \centering
  \includegraphics[scale=0.49]{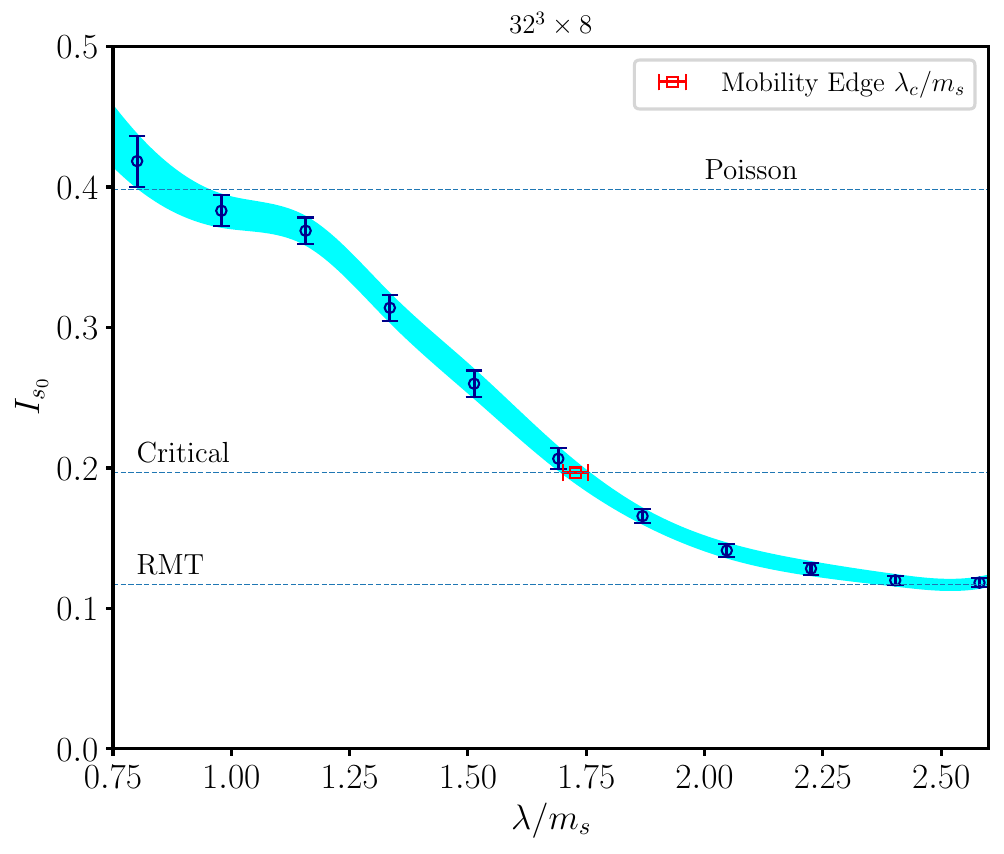}
  \includegraphics[scale=0.49]{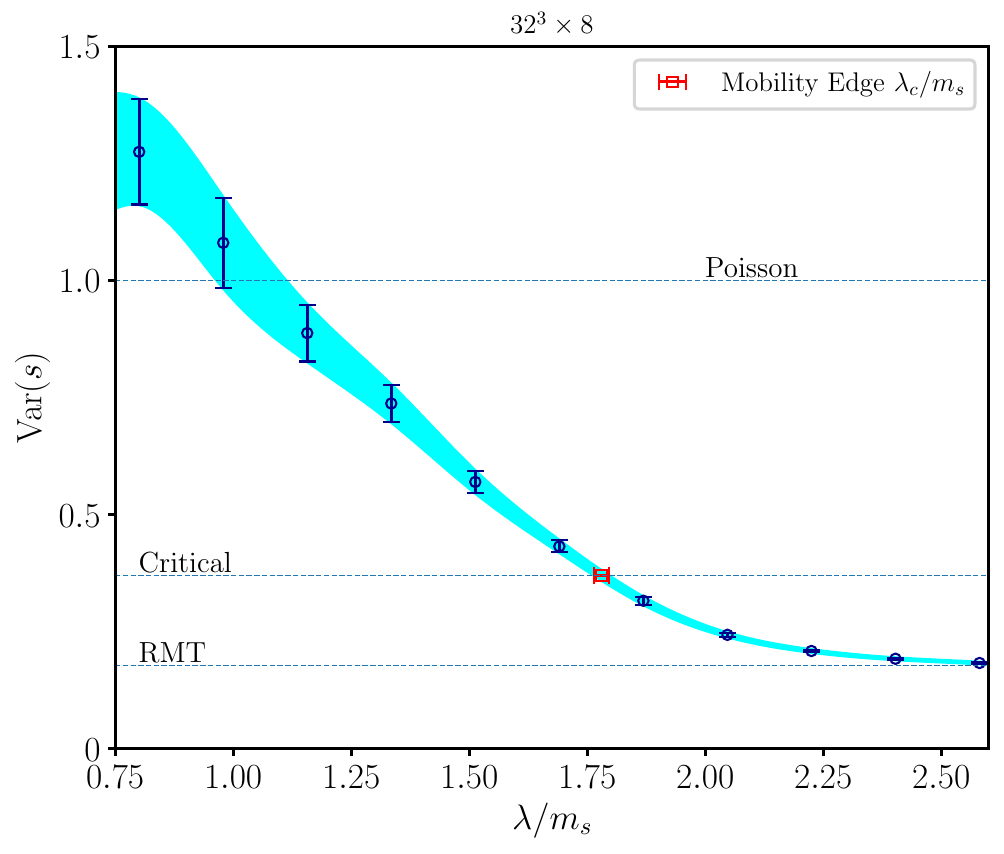}
  \caption{Determination of the mobility edge for our $32^3 \times 8$
    lattice from the dependence of $I_{s_0}$ (top panel) and $\Var(s)$
    (bottom panel) on the spectral region.}
  \label{fig:mob_edge_ex_Nt_8}
\end{figure}

The qualitative behavior of the dependence of
$\lambda_c^{\mathrm{stat}}/m_s$ on the lattice spacing at fixed aspect
ratio can be easily deduced from Fig.~\ref{fig:mob_edge_COMP}: as
$a\to 0$, both $I_{s_0}$ and $\Var(s)$ tend to increase throughout the
spectrum, as expected, leading to a larger estimate for the mobility
edge. In particular, $I_{s_0}$ overshoots the Poisson expectation at
the low end of the spectrum already on our coarsest lattices (even on
the $32^3\times 8$, where this happens outside of the spectral window
displayed in the plot), showing that taste degeneracy is already
distorting the spectral statistics there.  While this is not a problem
as it does not affect the region where the mobility edge actually is
(see below Sec.~\ref{sec:lc_cont_final}), as the lattice becomes finer
$I_{s_0}$ also overshoots its RMT expectation deeper in the bulk of
the spectrum, signaling that the effects of taste multiplets on
spectral statistics become important there, too.

The data show that the two coarsest lattices ($N_t=8,10$) give
compatible results for the spectral statistics in all spectral regions
in the bulk of the spectrum, and down to the first bin below the
mobility edge. 
Barring a conspiracy, this indicates two things: that the formation of
taste multiplets does not have significantly large effects in that
spectral region yet; and that, when this is the case, further lattice
artifacts, on top of those introduced by taste-degenerate multiplets,
are small when considering localization properties at fixed physical
volume. Clearly, there could still be finite-volume systematics, but
these would involve only effects unrelated to taste multiplets (such
as the localization length in the spectral region near the mobility 
edge being too large compared with the spatial size of the system), 
which are expected not to affect much the determination of the 
mobility edge via the matching with the critical statistics (see above 
in Sec.~\ref{sec:specstat}). Indeed, previous numerical results show that
for lattices of similar size and spacing one is already close to the
one-parameter scaling regime near the mobility
edge~\cite{Giordano:2013taa}. This shows that the determination of the
mobility edge from the critical spectral statistics is reliable on
these lattices.

The dependence of $\lambda_c^{\mathrm{stat}}/m_s$ on the aspect ratio
at fixed lattice spacing is visible instead in
Fig.~\ref{fig:mob_edge_COMP_Ns}, where $N_t=14$ and the spatial size
is varied. Here a larger aspect ratio drives $I_{s_0}$ and $\Var(s)$
down, as expected, so leading to an estimate for the mobility edge
that decreases with the spatial volume. Notice that the
scale-invariant nature of the mobility edge is still masked here by
the distortions of the unfolded level spacing distribution, and no
volume-independent crossing point of the various curves is
present. This shows that the effects of taste degeneracy are still
strong even on our largest lattice at $N_t=14$. This is in agreement
with the argument discussed in Sec.~\ref{sec:spec_taste}, as,
according to Eq.~\eqref{eq:dDest}, we would need an aspect ratio of
$LT\sim 9$ at this lattice spacing in order to have the same
$\delta/\Delta$ that we have for $N_t=8$ with $LT=4$. We then
expect that the plateau in $\lambda_c^{\mathrm{stat}}/m_s$ against
$LT$ is not visible yet, and would appear for an aspect ratio
roughly twice as big as our largest one.

In order to estimate $\lambda_c^{\mathrm{stat}}/m_s$ quantitatively,
we interpolated our numerical results for $I_{s_0}$ and $\Var(s)$ with
splines, defining an uncertainty band by interpolating the central
values augmented or reduced by the error. We then determined
$\lambda_c^{\mathrm{stat}}/m_s$ as the center of the interval where
the band crosses the critical value, and the corresponding error as
the half-width of this interval. This procedure is visualized in
Fig.~\ref{fig:mob_edge_ex_Nt_8} for our $32^3\times 8$ lattice. Our
final results are summarized in Tab.~\ref{tab:res_mob_edge}.

\begin{table}[!t]
  \begin{center}
    \begin{tabular}{|c|c|c|c|c|}
      \hline
      \multicolumn{5}{|c|}{}\\[-1em]
      \multicolumn{5}{|c|}{$T=230$ MeV, $\quad T/T_c=1.48$}\\
      \hline
      &&&&\\[-1em]
      $N_s$ & $N_t$ & $N_s/N_t$ & \makecell{$\lambda_c/m_s$\\(From $I_{s_0}$)} & \makecell{$\lambda_c/m_s$\\(From $\Var(s)$)} \\
      \hline
      &&&&\\[-1em]
      32 & 8 & 4 & 1.727(26)  & 1.780(15) \\
      \hline                                      
      &&&&\\[-1em]                                
      40 & 10 & 4 & 1.742(30) & 1.817(12) \\
      \hline                                      
      &&&&\\[-1em]                                
      48 & 12 & 4 & 1.865(23) & 1.979(21) \\
      \hline
      &&&&\\[-1em]
      48 & \multirow{3}{*}{14} & 3.4 & 2.124(61) & 2.235(26) \\
      56 &                     & 4   & 1.985(27) & 2.088(13) \\
      64 &                     & 4.6 & 1.889(36) & 1.955(20) \\
      \hline
      &&&&\\[-1em]
      64 & 16 & 4 & 2.112(43) & 2.199(39) \\
      \hline
    \end{tabular}
  \end{center}
  \caption{Summary of our results for the mobility edge in units of
    the strange quark mass, obtained by matching the behavior of the
    spectral statistics with the expected critical behavior.}
  \label{tab:res_mob_edge}
\end{table}

\subsection{Taste-degeneracy effects on the mobility edge in the thermodynamic limit}

\begin{figure}[!t]
  \centering
  \includegraphics[scale=0.46]{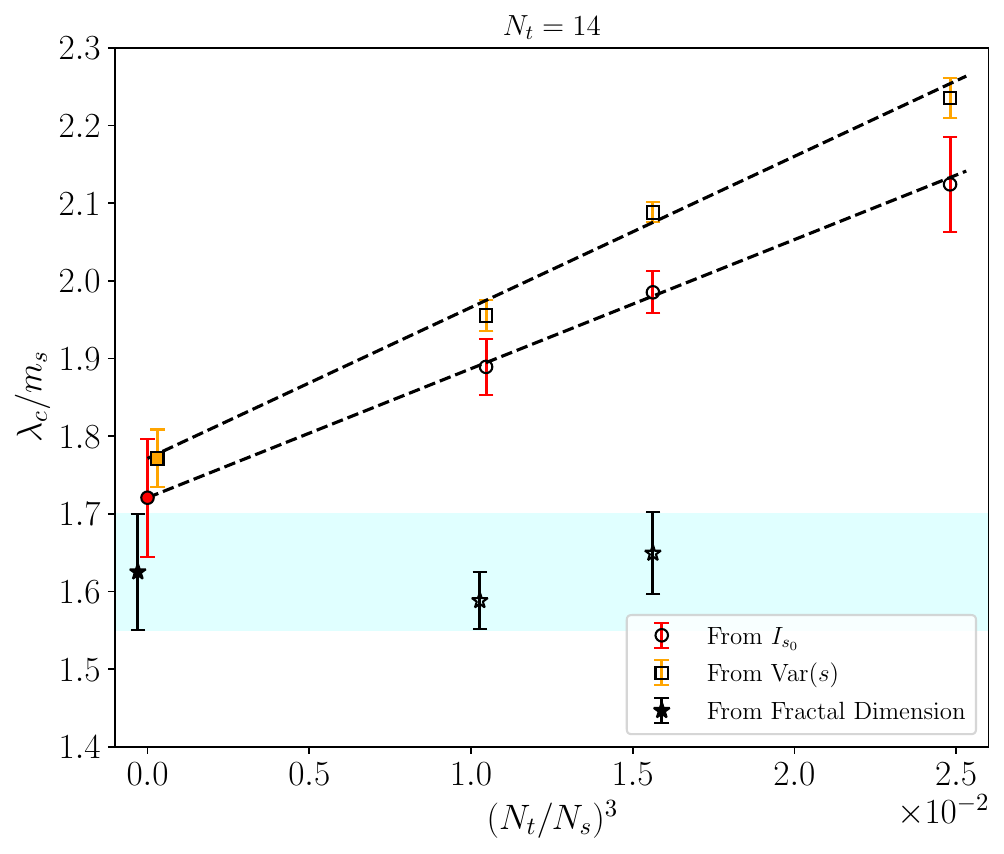}
  \caption{Dependence of our estimates for the mobility edge on the
    spatial size of the lattice, at fixed temporal extension
    $N_t=14$. The dashed lines represent thermodynamic extrapolations
    of $\lambda_c^{\mathrm{stat}}/m_s$ assuming leading linear
    corrections in the inverse spatial volume
    $(N_t/N_s)^3=1/(LT)^3$. The filled points at zero are the results
    of our extrapolations. Empty and filled starred points, and the
    shaded area, are our estimates of the mobility edge from the
    fractal dimension, see text for more details. Points have been
    slightly shifted horizontally to improve readability.}
  \label{fig:fse_mob_edge_Nt_14}
\end{figure}

\begin{figure}[!t]
\centering
\includegraphics[scale=0.49]{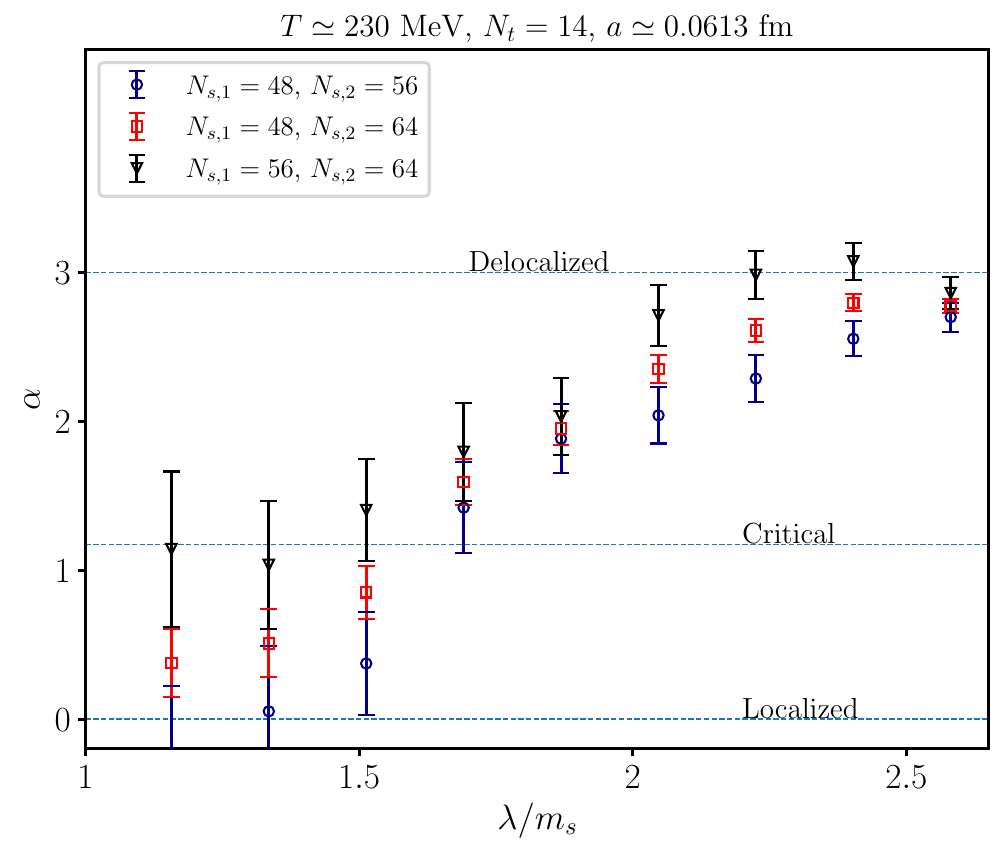}
\caption{Behavior of the fractal dimension as a function of the
  spectral bin computed for $N_t=14$ for several choices of the
  lattice sizes $N_{s,1}$ and $N_{s,2}$.}
\label{fig:frac_dim_COMP}
\end{figure}

\begin{figure}[!t]
  \centering
  \includegraphics[scale=0.49]{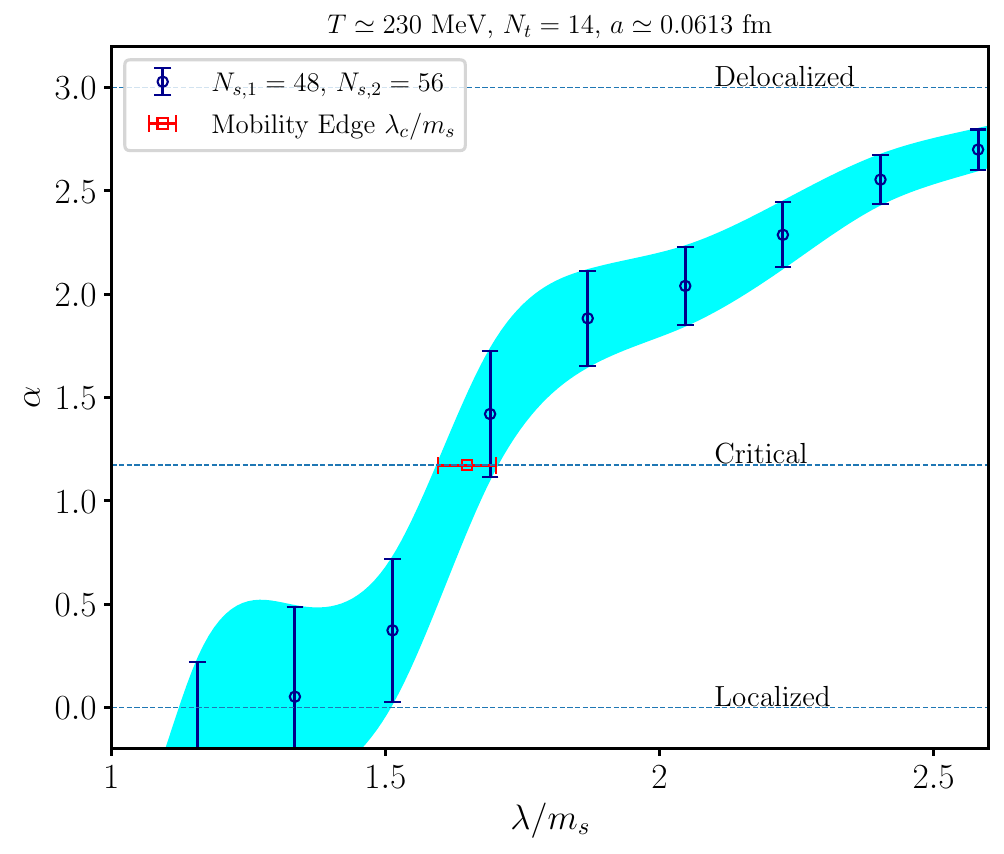}
  \caption{Determination of the mobility edge from the fractal dimension
    of eigenmodes, estimated using results from $48^3 \times 14$ and
    $56^3 \times 14$ lattices [see Eq.~\eqref{eq:alpha_num}].}
  \label{fig:fractal_dim_mob_edge_Nt_14}
\end{figure}

In order to show that the effects of taste degeneracy become
irrelevant for the determination of the mobility edge with our method
when taking the thermodynamic limit at fixed lattice spacing, we have
compared the $N_s\to\infty$ extrapolation of our $N_t=14$ data with an
independent determination of $\lambda_c$ based directly on the
properties of the eigenvectors. Distortive effects on the statistics are 
due to approximate taste degeneracy, which is visible when
$(a\Lambda)^2 < 1/(V\Lambda^3)$.  We then expect finite-size
corrections to $\lambda_c^{\mathrm{stat}}/m_s$ of order $O(1/N_s^3)$,
up to lattice sizes where the effects of taste degeneracy
disappear.  As these are expected to be about an order of magnitude
larger than our largest lattices, we can ignore the onset of the
plateau in our extrapolation in $(N_t/N_s)^3$.
The infinite-volume extrapolation is
shown in Fig.~\ref{fig:fse_mob_edge_Nt_14}; the corresponding results
for the mobility edge are:
\beq
\label{eq:therm_lim_Is0}
\f{\lambda_c}{m_s} &= 1.721(76), \qquad &(\text{from } I_{s_0}),\\
\label{eq:therm_lim_var}
\f{\lambda_c}{m_s} &= 1.771(37), \qquad &(\text{from } \Var(s)).
\eeq
To determine $\lambda_c$ directly from the eigenvectors, we estimated
the local fractal dimension numerically as
\begin{equation}
  \label{eq:alpha_num}
  \alpha_{\mathrm{num}}(\lambda) = 3 + \f{\log \left(
      \overline{\mathrm{PR}}(\lambda,N_{s2})/
      \overline{\mathrm{PR}}(\lambda,N_{s1})\right)}{\log 
    ( N_{s2}/N_{s1})}\,.
\end{equation}
Our results are shown in Fig.~\ref{fig:frac_dim_COMP}.  We then looked
for the point in the spectrum where $\alpha_{\mathrm{num}}$ takes its
critical value $\alpha_* = 1.173_{-26}^{+32}$~\cite{Ujfalusi:2015mxa},
using a spline interpolation of the numerical data. This is visualized
in Fig.~\ref{fig:fractal_dim_mob_edge_Nt_14}. To quote a final
  value, we considered the results obtained with the pairs
  $(N_{s1},N_{s2})=(48,56)$ and $(48,64)$, and took as our final
  estimate a symmetric confidence band including both error bars. The
two values are compatible within errors (see
Fig.~\ref{fig:fse_mob_edge_Nt_14}), showing that finite-size effects
on this estimate of the mobility edge are reasonably small. As a
further check, we verified that the pair $(N_{s1},N_{s2})=(56,64)$
gave compatible results, although within much larger statistical
errors. In the end, we obtain:
\beq
\label{eq:mob_edge_fract_dim}
\f{\lambda_c}{m_s} = 1.625(75), \quad \, (\text{from fractal
  dimension}).  \eeq
This is in good agreement with the thermodynamic extrapolation of our
estimates from spectral statistics, Eq.~\eqref{eq:therm_lim_Is0}, see
Fig.~\ref{fig:fse_mob_edge_Nt_14}. This shows that these are indeed
converging to the actual position of the mobility edge, as we
  argued in Sec.~\ref{sec:spec_taste}.  Moreover, this further
confirms that finite-size effects on the estimate based on the
critical fractal dimension are reasonably small.

\subsection{Dependence of the mobility edge on the lattice spacing
  and exchange of continuum and thermodynamic limits}
\label{sec:lc_cont}

\begin{figure}[!t]
\centering
\includegraphics[scale=0.49]{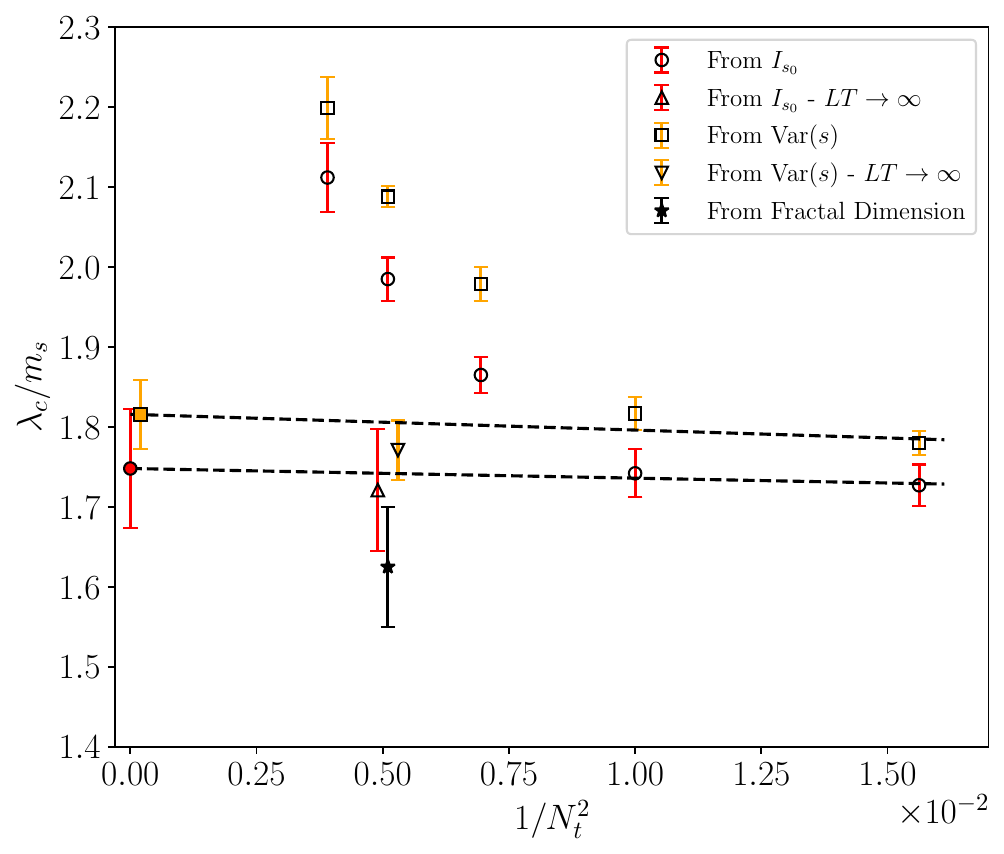}
\caption{Dependence of our estimates for the renormalized mobility
  edge as a function of the lattice spacing at fixed temperature
  $T=230\,\mathrm{MeV}$. The dashed lines represent continuum
  extrapolations of $\lambda_c/m_s$ assuming leading linear
  corrections in $(1/N_t)^2=(aT)^2$, see text for details. Points have
  been slightly shifted horizontally to improve readability.}
\label{fig:mob_edge_cont_lim}
\end{figure}

We finally summarize the results of the previous two subsections in
Fig.~\ref{fig:mob_edge_cont_lim}, showing together all our
determinations of $\lambda_c/m_s$. It is clear that our two coarsest
lattices give compatible estimates via $\lambda_c^{\mathrm{stat}}/m_s$,
while on finer lattices with the same aspect ratio these estimates
rapidly deviate.  On the other hand, the infinite-volume extrapolation
at $N_t=14$ gives again results compatible with our coarsest lattices,
as well as with the determination from the local fractal dimension.

Before attempting an extrapolation of $\lambda_c/m_s$ to the
continuum, it is important to check that our numerical results are in 
agreement with our theoretical argument developed in Sec.~\ref{sec:spec_taste_cont}
that the thermodynamic and continuum limits can be exchanged.
While a direct check is not possible with the available data, indirect
support is provided by our results for the estimate based on
$\alpha(\lambda)$. Being unaffected by taste degeneracy, this gives us
a valid estimate of the position of the mobility edge at both finite
aspect ratio and finite spacing.  We have already pointed out above
that this direct estimate of the mobility edge does not depend strongly on
the aspect ratios employed, and agrees with the indirect estimate from
spectral statistics extrapolated to infinite volume. Moreover, our
results obtained at $N_t=14$ are not very different from the estimates
obtained using spectral statistics on our coarsest lattices
$N_t=8,10$, where taste-degeneracy effects on spectral statistics are
under control and estimates based on critical statistics should
accurately reflect the position of the mobility edge (within our
numerical accuracy).  
These observations suggest that the true position of
the mobility edge, as intrinsically defined by 
the properties of the eigenvectors, does not depend
strongly neither on the aspect ratios nor on the lattice spacing.
This is in agreement with our argument that the continuum and thermodynamic
limit commute for what concerns the mobility edge.
We can then use indirect estimates of the mobility edge from spectral statistics 
extrapolated to the thermodynamic limit, followed by the continuum limit, to obtain the correct
value of the mobility edge in the continuum theory.

Further numerical support to the correctness of our
theoretical argument can be obtained if we manage to remove the
effects of taste degeneracy from the spectrum.  For a suitably
``taste-symmetrized'' spectrum, one expects a more accurate
correspondence between localization properties of the eigenvectors
and spectral statistics already at finite volume.  Agreement of the
resulting estimates for the mobility edge with the ones obtained by
using the spectrum without further processing and taking the
thermodynamic limit, would provide further support to the
correctness of our procedure. Clearly, the taste-symmetrization of
the spectrum is an ambiguous procedure, even if one could correctly
identify the eigenvalue multiplets, which is not easy when the
spacing is too coarse; even on our finest lattices, only doublets
are visible.  This ambiguity leads to hardly quantifiable systematic
effects, that make this approach less reliable for a controlled
continuum extrapolation.

\begin{figure}[!t]
  \centering
  \includegraphics[scale=0.49]{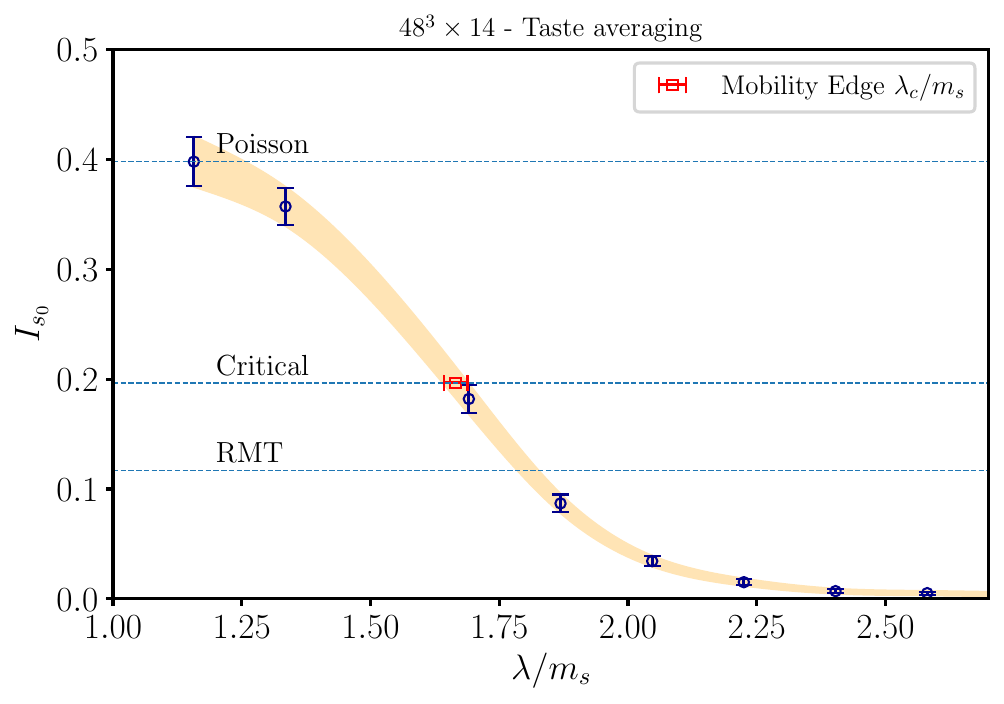}
  \includegraphics[scale=0.49]{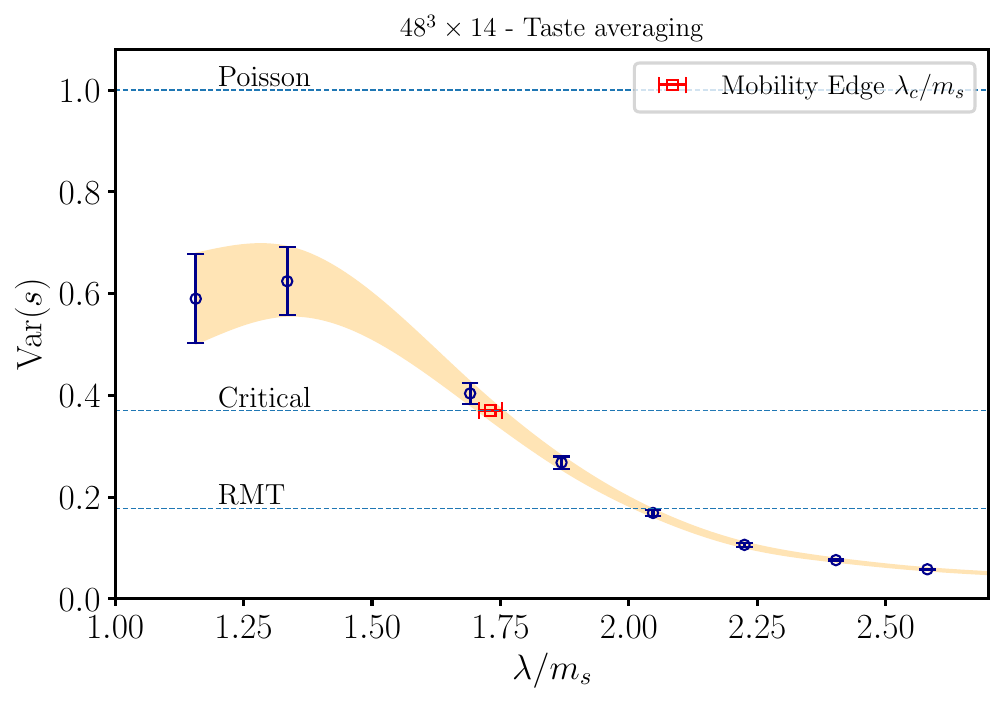}
  \caption{Determination of the mobility edge from the
      spectral statistics of the spectrum after taste averaging 
      (see text for details), estimated using results from $48^3 \times 14$ lattices.}
  \label{fig:mob_edge_Nt_14_doub}
\end{figure}

To obtain a heuristic estimate we have then opted for the simplest
approach, namely replacing pairs of neighboring eigenvalues with
their arithmetic average. This has a number of obvious theoretical
issues, first of all the likely misidentification of at least part of the
doublets. This problem is particularly severe in the bulk of the
spectrum, where doublet misidentification is almost certain, leading
to interfering with the relevant spectral correlations and possibly
to destroying them completely. Nonetheless, one expects that for finer 
and finer lattices at fixed aspect ratio this procedure should work better 
and better (and worse and worse on bigger and bigger lattices at fixed lattice spacing), 
and eventually provide the correct result in the continuum limit 
(if one extends it to quartets when these show up); and one can hope that 
for sufficiently fine but still manageable lattices it works reasonably well 
up to and slightly above the mobility edge.
%
We have then averaged the spectrum as discussed above, and performed
on it the same analysis as with the untreated spectrum, looking for
the point where spectral statistics match their critical value.  We
restricted the spectrum to regions where the unfolded symmetrized
level spacings satisfy $\la s\ra \simeq 1$, as a minimum requirement
for the reliability of the procedure. This is illustrated in
Fig.~\ref{fig:mob_edge_Nt_14_doub}.

\begin{figure}[!t]
  \centering
  \includegraphics[scale=0.49]{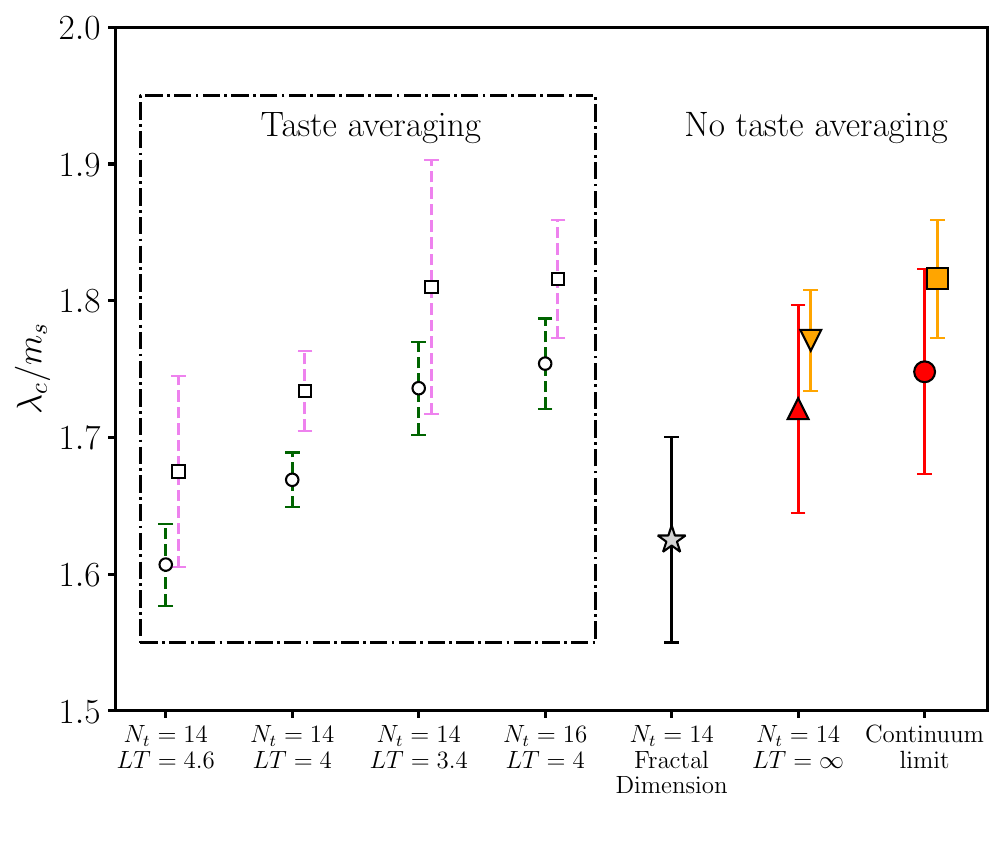}
  \caption{Estimates for the mobility edge
      from the spectral statistics of the spectrum after 
      taste averaging (see text for details), obtained using results from our 
      $N_t=14,16$ lattices. 
      For comparison, we show also our results from the
      fractal dimension and from standard spectral statistics 
      extrapolated to the thermodynamic limit on $N_t=14$ lattices, 
      and our continuum extrapolation 
      based on standard spectral statistics (see Sec.~\ref{sec:lc_cont_final}).}
  \label{fig:comp_mob_edge_Nt_14_doub}
\end{figure}

With this long list of caveats in mind, we show our results in
Figs.~\ref{fig:mob_edge_Nt_14_doub} and
\ref{fig:comp_mob_edge_Nt_14_doub}.  As expected, RMT correlations in the
bulk of the spectrum are lost. However, in spite of our rather crude
approach, the agreement between estimates based on the 
taste-averaged and untreated spectra is surprisingly good already for our $N_t=14$
lattices. While uncertainties are hard to quantify, we believe that these
results provide strong numerical support to our theoretical argument
for the exchangeability of continuum and thermodynamic limits.

For future utility, we report also the results obtained with the
taste-averaging procedure on our finest, $N_t=16$ lattice, i.e.,
$\lambda_c/m_s=1.754(33)$ 
from $I_{s_0}$ and
$\lambda_c/m_s=1.766(54)$ 
from $\Var(s)$.

\subsection{Continuum extrapolation}
\label{sec:lc_cont_final}

In the light of this discussion, we have performed an extrapolation to
the continuum limit using our estimates of $\lambda_c/m_s$ from
spectral statistics for $N_t=8,10$ and $N_t=14$, where for $N_t=14$ we
have used the results extrapolated first toward the thermodynamic
limit. For the continuum extrapolation we have assumed the
behavior $\f{\lambda_c}{m_s}= \f{\lambda_c}{m_s}\big|_{\mathrm{cont}} +
C/N_t^2$, based on our theoretical expectations discussed in
Sec.~\ref{sec:spec_taste_cont}.  The results obtained from the two
observables $I_{s_0}$ and $\Var(s)$ are:
\beq
\label{eq:cont_lim_Is0}
\f{\lambda_c}{m_s}\bigg|_{\mathrm{cont}} &= 1.748(75), \qquad &(\text{from } I_{s_0}),\\
\label{eq:cont_lim_var}
\f{\lambda_c}{m_s}\bigg|_{\mathrm{cont}} &= 1.816(43),
\qquad &(\text{from } \Var(s)),
\eeq
perfectly compatible within errors. We note that, while
affected by uncontrolled theoretical uncertainties, the results
obtained with our taste-averaging procedure on our finest,
$N_t=16$ lattice, reported above at the end of
Sec.~\ref{sec:lc_cont}, are in excellent agreement with these
estimates (see Fig.~\ref{fig:comp_mob_edge_Nt_14_doub}).

To quote a final number, we did a weighted average of our two estimates and their statistical errors, 
and used their difference as an estimate of systematic finite-size effects. We find:
\begin{align}
\label{eq:cont_lim_ave}
    \f{\lambda_c}{m_s}\bigg|_{\mathrm{cont}} &= 
    1.799 (51)_{\mathrm{stat}} (68)_{\mathrm{syst}}, 
   \\
   \label{eq:cont_lim_ave_comb}
   &= 
    1.799 (85)_{\mathrm{combined}}.
\end{align}
To our knowledge, this result is
the first fully controlled extrapolation of $\lambda_c/m_s$ to the
continuum.

\section{Conclusions}
\label{sec:concl}

Localization of the low-lying Dirac eigenmodes in the high-temperature
phase of QCD and other gauge theories~\cite{Garcia-Garcia:2006vlk,
  Kovacs:2012zq,Giordano:2013taa,Dick:2015twa,Ujfalusi:2015nha,
  Cossu:2016scb,Holicki:2018sms,Kehr:2023wrs,Meng:2023nxf,
  Giordano:2021qav,Gockeler:2001hr,Gattringer:2001ia,Gavai:2008xe,
  Kovacs:2009zj,Kovacs:2010wx,Bruckmann:2011cc, Kovacs:2017uiz,
  Giordano:2019pvc,Vig:2020pgq,Bonati:2020lal,Alexandru:2021pap,
  Alexandru:2021xoi,Baranka:2021san,Baranka:2022dib,Alexandru:2023xho,
  Giordano:2016nuu,Cardinali:2021fpu,Baranka:2023ani} is closely
related to the change in the confining properties of these theories
taking place at the transition~\cite{Bruckmann:2011cc,Giordano:2015vla,Giordano:2016cjs,
  Giordano:2016vhx,Giordano:2021qav,Baranka:2022dib,Kovacs:2017uiz,
  Giordano:2019pvc,Vig:2020pgq,Bonati:2020lal,Baranka:2021san,
  Baranka:2022dib,Giordano:2016nuu,Cardinali:2021fpu}, and its study
can lead to a better understanding of the mechanism behind
deconfinement, and of its relation with chiral symmetry
restoration.

The strong connection between deconfinement and localization is
exemplified by the fact that the mobility edge, i.e., the point in the
spectrum separating localized and delocalized modes in the spectrum,
decreases as one approaches the pseudocritical temperature from above,
and vanishes in the crossover range. In theories with a genuine
deconfinement transition, this takes place exactly at the critical
point~\cite{Kovacs:2017uiz,Giordano:2019pvc,Vig:2020pgq,
  Bonati:2020lal,Baranka:2021san,Baranka:2022dib,Giordano:2016nuu,
  Cardinali:2021fpu}. A more accurate quantitative determination of
the ``geometric'' transition temperature where the mobility edge
vanishes and localization disappears, and so of the tightness of the
connection between localization and deconfinement, requires the full
control of systematic effects, including finite volume and,
especially, finite spacing effects.

Most of the numerical studies of localization combine the relation
between the localization properties of eigenmodes and the statistical
properties of the corresponding
eigenvalues~\cite{altshuler1986repulsion} with the use of the
staggered discretization of the Dirac
operator~\cite{Garcia-Garcia:2006vlk,Kovacs:2012zq,Giordano:2013taa,
  Ujfalusi:2015nha,Bonati:2020lal,Cardinali:2021fpu,Kovacs:2010wx,
  Bruckmann:2011cc,Kovacs:2017uiz,Giordano:2019pvc,Baranka:2021san,
  Baranka:2022dib,Giordano:2016nuu,Baranka:2023ani}. However, this
approach faces serious technical problems due to the restoration of
taste symmetry in the continuum limit: as the lattice becomes finer
the spectrum tends to organize in nearly-degenerate multiplets of
eigenmodes, which in turn distort the spectral statistics away from
their expected universal behavior, leading to hard-to-control
systematic effects in the determination of the mobility edge.

In this paper we have carried out a dedicated study of the effects of
taste degeneracy on the statistical properties of the staggered
spectrum in high-temperature lattice QCD. We focused in particular on
how these affect the numerical determination of the mobility edge, and
how these effects change as the lattice spacing is reduced, or the
aspect ratio is increased, with the main goal of providing a
controlled continuum limit of the mobility edge.
To this end, we studied in detail the possibility of exchanging the 
order of the continuum and thermodynamic limits, arguing that 
it applies to the study of localization properties and the
determination of the mobility edge.

Our findings are in line with theoretical expectations, with a
systematic overestimation of the mobility edge 
using spectral statistics on finer lattices,
where the effects of taste degeneracy become sizeable also in the bulk
of the spectrum. For larger aspect ratios at fixed lattice spacing
these effects are reduced and the overestimation of the mobility edge
is mitigated. In the thermodynamic limit, this estimate tends to the
correct value of the mobility edge, obtained independently by studying
the fractal dimension of the eigenvectors. Moreover, it 
agrees 
with 
finite-volume estimates obtained from spectral statistics using a suitably 
taste-symmetrized spectrum, for which the effects of taste degeneracy should be
reduced. This supports our expectation that continuum and
thermodynamic limits can be exchanged 
when studying the
localization properties of Dirac eigenmodes.

From a practical perspective, the most important result of this
analysis is that the infinite-volume extrapolation of the mobility
edge on a finer lattice is in good agreement with the mobility edge
found on coarser lattices, where taste-degeneracy effects do not reach
into the bulk of the spectrum and a reliable estimate can be obtained
from spectral statistics already at lower aspect ratios. This shows
that accurate values for the mobility edge can be obtained on
relatively coarse lattices for reasonable aspect ratios using spectral
statistics, which is the ideal combination from the numerical point of
view.

Moreover, our findings allowed us to perform the first, fully
controlled extrapolation of the mobility edge to the continuum
limit. This confirms the theoretical
expectation~\cite{Kovacs:2012zq,Giordano:2022ghy} that the mobility
edge in units of the quark mass is a renormalized quantity with
physical meaning, and not only a lattice artifact. It also lends
support to the numerical evidence for this fact presented in
Refs.~\cite{Kovacs:2012zq,Cardinali:2021fpu}. There it was shown that
the mobility edge depends only mildly on the lattice spacing.  In the
light of our results, this is because the calculations of
Refs.~\cite{Kovacs:2012zq,Cardinali:2021fpu} use relatively coarse
lattices, where taste-degeneracy effects on 
estimates of the mobility edge based on spectral statistics are
negligible. One might have wondered if the situation could have
changed on finer lattices; our results show that this mild dependence
would still show on finer lattices, provided one extrapolated first to
the thermodynamic limit. (Incidentally, the extrapolation in $T$ shown
in Ref.~\cite{Kovacs:2012zq} is in good qualitative agreement with our
result.)

In conclusion, we have shown that one can use staggered fermions
efficiently and reliably to study the mobility edge in the Dirac
spectrum of high-temperature lattice gauge theories, provided that the
aspect ratio is sufficiently large to avoid sizeable taste-degeneracy
effects in the relevant region of the spectrum. 
The possibility to bypass these effects through this procedure
is due to the possibility of exchanging the continuum and
thermodynamic limits when studying localization properties.
Moreover, we have numerically demonstrated that the mobility edge in
units of the quark mass is a renormalized quantity, in agreement with
theoretical expectations; and that it depends only mildly on the
lattice spacing.

It would be interesting to understand if and how one could avoid the
well known problems of the staggered discretization (i.e., lack of
good chiral properties, difficult relation with topology) by relating
the staggered and the overlap spectrum on the same gauge
configurations.  In this context, the mobility edge could be used in
two different ways. On the one hand, renormalizing the overlap
spectrum (after matching a suitable observable to find the
renormalization constant) would allow one to obtain another estimate
for $\lambda_c/m_s$ from spectral statistics unaffected by taste
degeneracy, to be compared with the one obtained here after
extrapolating to infinite volume. On the other hand, thanks to its
mild dependence on the lattice spacing the mobility edge itself could
be efficiently used to match the staggered and overlap spectra.

\begin{acknowledgments}
  It is a pleasure to thank Massimo D'Elia and Tam{\'a}s G.~Kov{\'a}cs
  for useful discussions and for a careful reading of the manuscript,
  and Giuseppe Clemente for help in setting up the diagonalization
  code.  The work of C.~B.~is supported by the Spanish Research Agency
  (Agencia Estatal de Investigaci\'on) through the grant IFT Centro de
  Excelencia Severo Ochoa CEX2020-001007-S and, partially, by grant
  PID2021-127526NB-I00, both funded by
  MCIN/AEI/10.13039/501100011033. C.~B.~also acknowledges support from
  the project H2020-MSCAITN-2018-813942 (EuroPLEx) and the EU Horizon
  2020 research and innovation programme, STRONG-2020 project, under
  grant agreement No 824093. M.~G.~was partially supported by the
  NKFIH grant KKP-126769, and by the NKFIH excellence grant
  TKP2021-NKTA-64.  Numerical calculations have been performed on the
  \texttt{Finisterrae~III} cluster at CESGA (Centro de
  Supercomputaci\'on de Galicia). We also acknowledge the use of gauge
  configurations and Dirac spectra previously obtained using the
  \texttt{Marconi100} machine at Cineca, based on the agreement
  between INFN and Cineca (under projects INF19\_npqcd, INF20\_npqcd
  and INF21\_npqcd).
\end{acknowledgments}

\bibliographystyle{apsrev4-2}
\bibliography{references_taste}

\end{document}